\documentclass[11pt, a4paper]{article}

\usepackage{amsmath, amssymb}
\usepackage{graphicx}
\usepackage{fancybox}
\usepackage{lipsum}
\PassOptionsToPackage{hyphens}{url}
\usepackage{hyperref}
\usepackage{apacite}
\usepackage{color}

\begin{document}

\title{Timescales: the choreography of classical and unconventional computing}

\author{Herbert Jaeger\thanks{corresponding author} \\\emph{Bernoulli Institute and CogniGron} \\\emph{University of Groningen} \and Francky Catthoor \\ \emph{IMEC Belgium in Leuven} \\ \emph{KU Leuven} }

\date{January 18, 2023}

\maketitle

\begin{abstract}
  Tasks that one wishes to have done by a computer often come with an
  assortment of conditions that relate to timescales in one way or the
  other. For instance, the processing must terminate within a given
  time limit; or a signal processing computer must integrate input
  information across several timescales; or a robot motor controller
  must react to sensor signals with a short latency. For classical
  digital computing machines such requirements pose no fundamental
  problems as long as the physical clock rate is fast enough for the
  fastest relevant timescales. However, when digital microchips become
  scaled down into the few-nanometer range where quantum noise and
  device mismatch become unavoidable, or when the digital computing
  paradigm is altogether abandoned in analog neuromorphic systems or
  other unconventional hardware bases, temporal task conditions are
  more difficult to relate to the physical hardware dynamics, and
  instructive literature is scarce. --- Here we explore the relations
  between task-defined timescale conditions and physical hardware
  timescales in some depth and breadth. The article has two main
  parts. In the first part we develop an abstract model of a generic
  computational system that admits a unified discussion of various
  kinds of computational timescales. This model is general enough to
  cover digital computing systems as well as analog neuromorphic or
  other ``unconventional'' ones. We identify four major types of
  ``timescales'' which require separate considerations: causal
  physical timescales which are formally expressed by time constants;
  timescales of phenomenal change which characterize the ``speed'' of
  how something changes in time; timescales of reactivity which
  describe how fast a computing system can react to incoming trigger
  information; and memory timescales. In the second part we survey
  twenty known computational mechanisms that can be used to obtain
  desired task-related timescale characteristics from the
  physical givens of the available hardware.\\

  \emph{This report is a substantially revised and extended version of a
deliverable written for the EU H2020 project ``MemScales'' (grant
number 871371 \url{https://memscales.eu})}

  \end{abstract}

  \tableofcontents

  \newpage

\section{Introduction}
\label{secIntro}
Digital information processing technologies are increasingly
challenged by conflicts between exploding data volumes, ubiquitous
demands for internet services, and intelligent processing complexity
on the one hand, versus energy consumption and miniaturization
barriers on the other. This ``end of Moore's law'' theme with its many
facets has become recited so often that we do not need to expand on it
here. Besides quantum computing -- which we explicitly do not consider
in this article -- we perceive three main strategies to meet this
challenge:
\begin{itemize}
\item miniaturizing digital microchips to the very limits with new
  \emph{deep sub-micron scaled} device technologies, finding new
  computational methods to cope with increasing
  physical device mismatch and stochasticity \cite{HoriguchiTokei20},
    \item ``learn from the brain'' and explore \emph{neuromorphic
        computing} paradigms and hardware bases \cite{naturecomm_qa19},
    \item find ways to ``exploit the physics of the materials directly'' \cite{Zauner05} in  \emph{unconventional computing} paradigms (also known under other names like in-materio computing, physical computing, etc.).
\end{itemize}
These research lines venture into novel materials, devices and
physical phenomena. A number of challenges arise
in some or all of them. They are more fundamental than mere technical
obstacles which will just take a good deal of effort, time and funding
to be solved:

\begin{itemize}
\item Device mismatch, aging, and physical degradation mandates
  compensatory computational mechanisms for fabrication-time or
  runtime calibration, dynamical stabilization, or error recovery.
\item Fabrication variability makes it impossible to produce
  functionally equivalent clones of analog neuromorphic,
  unconventional, and to some degree also deep sub-micron
  microchips. Every chip may need individual formation or training.
\item Not all functionally relevant physical quantities and processes
  can be physically measured on-board. Spatially continuous nonlinear
  substrates, which sometimes are considered in unconventional
  computing, may even be mathematically and functionally
  infinite-dimensional. In deeply scaled digital computing restricted
  physical observability hinders the development and testing of
  accurate simulation models and the testing and debugging of
  algorithmic procedures \cite{Abrahametal02, Schat09}.
\item Neuromorphic and
  unconventional computing systems often cannot be programmed in the classical
  sense but will have to be configured, trained, or evolved, which
  requires new views on how to use such systems for which tasks.
\item Thermal or quantum noise and temperature
  sensitivity lead to low, variable, or even ill-defined
  numerical precision, which requires the development of new
  mathematical formats for representing or quantifying information.
\item Many neuromorphic and unconventional systems will age or become
  physically transformed during their lifetime by learning and
  adaptation, which implies that they cannot be ``re-started'' or
  ``re-set'' to some well-defined initial state in order to execute new ``runs''.
\end{itemize}

Such difficulties challenge our fundamental conception of what
``computing'' is. Turing computability and symbolic, deterministic
algorithms may turn out to be incompatible with such phenomena.
Generalized concepts of ``computing'' have been sought since long in
several disciplines, especially in theoretical biology and
neuroscience, theoretical physics and various unconventional
niches in computer science.  It is maybe more than a
historical curiosity that Turing himself in his last works laid the
foundations for self-organized pattern formation in biological systems
\cite{Turing52}, addressing a family of phenomena that has later been
recruited as a basis for unconventional or brain-like kinds of
``computing'' \cite{Adamatzky11, LinsSchoener14}. The quest for a
generalized theory of ``computing'' has recently gained fresh momentum
\cite{Adamatzky17ab, StepneyRasmussenAmos18,Jaeger21a}.

In this article we address one of the big challenges for deep
sub-micron, neuromorphic and unconventional
computing which we did not yet mention in the listing above: coming to
terms with multiple timescales. We call this the \emph{computational
  timescale extension} (CTE) challenge:

\vspace{0.5cm}

\fbox{\parbox{13cm}{\emph{Given that a computing system can physically
      realize only a limited range of physical timescales while many
      tasks need a wider range, how can the physically available
      timescales be extended by purely computational mechanisms?}}}
\vspace{0.5cm} 
     
Classical digital technologies have a unique option to escape from the
CTE problem, provided that the clock speed is faster than the fastest
timescale required by the task. In such systems one can always stably
store interim information structures for times as long as needed.
Today's (and tomorrow's) deep-submicron technologies enable device
delays that approach 1 psec and component delays that are well below 1
nsec. These extremely small time constants potentially enable the
execution of computational models with very short timescales. However,
the very short timescales on highly miniaturized and hence densely
packed microchips come with their own new challenges. Different
voltage and clock frequency ``islands'' must be dynamically controlled
and overall meet the throughput and latency requirements of the entire
system. This requires a control process overhead which is apt to
diminish or cancel the efficiency gains of potentially very high
processing speeds.  New methods are being proposed for ultra-scaled
digital microchips \cite{Noltsisetal19} to remove that overhead and
make fuller use of the fast processing available at the physical level
while retaining timing guarantees. The timing challenges that are
connected with deeply scaled digital microchips have some surprising
connections with challenges in non-digital computing. 
While the main focus of this report is on
non-digital computing technologies, we will also include in our
discussion those aspects of deeply
scaled digital systems which are akin to characteristics of analog
neuromorphic and other unconventional systems.

This report has two main sections, followed by a summary and
discussion. The first main Section \ref{secTheory} investigates CTE
from a methodological angle, clarifying the concept in the first
place. The second main Section \ref{secZoo} surveys CTE mechanisms
that we are aware of at this time.

\section{Theoretical considerations}\label{secTheory}

Computational timescale extension is an umbrella
term for a number of phenomena and techniques. Before we can discuss
CTE in a meaningful way we must work out our understanding of the
underlying conceptions of ``computing'' and ``timescales'' and also what
we mean by ``extension''. In this section we sharpen and differentiate
our understanding of the CTE challenge.

\subsection{Different physical systems, different computing paradigms,
  common challenges}

Let us first describe in more detail the three domains where
we find that the CTE challenges arise.

\subsubsection{Deeply scaled digital systems}\label{subsecDeeplyScaled}

Here is a simplified view of an ideal, classical digital
computing system. Its hardware is a network of small subsystems that
each serve an elementary computational function (most importantly
circuits for Boolean gates and non-volatile bistable memory
devices). These subsystem are connected by complex wire networks. Each
subsystem generates a binary output voltage if its input is a binary
voltage vector. Voltages are switched under the control of a global
clock cycle that is synchronously sent to all subsystems and whose
period length is long compared to the time that each subsystem needs
to stabilize its new output voltage; the overall electrodynamics can
thus be reliably abstracted to a Boolean network whose gates all
recompute their outputs synchronously (note that a memory device can
be realized by a Boolean circuit with self-feedback). Such ideal
digital hardware systems can be designed to yield the overall input-
to output functionality of any Turing machine, from simple binary
adders to universal programmable computers.

Ever increasing demands on data throughput rates have been addressed
on all levels from device technologies to computer and network
architectures. On the microchip technology side, the main answer is
miniaturization. The evolutionary pressure to minimize the size of
transistors and maximize their area density has since long eroded the
ideal picture of a digital computing system drawn above
\cite{SylvesterSylvester98, SylvesterSylvester99}. As device sizes
shrink and wiring becomes denser, a host of technological and
conceptual challenges rise their heads. They include, among many
other, variable and unpredictable signal travel delays, wire
crosstalk, conflicts between transistor miniaturization versus
response latency and leakage currents, device mismatch from
fabrication and during usage (including aging and external
disturbances like highly-energetic particles), and more such
detrimental physical phenomena. A rich literature is concerned with
methods to identify, monitor, control and compensate such challenging
effects, treating aspects like classification and testing of defects
\cite{Abrahametal02, Schat09}, mitigation of chip-lifetime timing degradation
\cite{Rodopoulosetal15, Weckxetal15, Jiangetal21}, or characterizing and
optimizing quantization losses \cite{Menardetal19}

The smaller digital devices and wires become, the more they exhibit
the variability and unreliability that we find in biological neurons
and synapses as well as in many analog neuromorphic and unconventional
nanoscale systems. The main strategy to cope with this in digital
nanoelectronics is to optimize the physical construction of
microchips, aiming at minimizing the impact of these effects such that
the final chip still behaves as ``classically'' as possible. This forces
digital microchip engineers to experiment with ever more intricate
designs for device and wiring shaping and spatial arrangement and to
explore new materials and fabrication technologies
\cite{HoriguchiTokei20}. But besides this preferred strategy of
finding physical hardware solutions, also software based approaches
are being explored. The idea here is to insert a ``nanoprogramming''
layer between the hardware and what normally is the lowest software
level (Instruction Set Architecture, ISA). This layer is hidden from
end-users, shielding them against the physical hardware variability
and unpredictability and provides them with a ``classical'' programming
interface to the microchip. This elementary software layer would have
to realize functionalities which are also needed in neuromorphic and
unconventional computing, for instance error checking and recovery,
exploiting redundancy, dynamical stabilization and calibration, or
synchronization and dynamical cycle time adaptation of the many local
physical clocks operating on the chip. The latter connects digital
nanoscale electronics to the theme of our report: computational
management of timescales.

\subsubsection{Neuromorphic computing}

``Neuromorphic computing'' is a cover term for a range of approaches in
an interdisciplinary landscape with overlaps to machine learning,
signals and systems, computational neuroscience, materials physics and
devices and microchip technologies. From an eagle's perspective, the
common denominator is to depart from the conception of ``computing'' as
the execution of symbolic algorithms, and instead view ``computing'' as
distributed, non-symbolic, ``brain-like'' information processing. Due to
a number of reasons --- the two primary ones being the successes of
deep neural networks and the promises of spiking neural dynamics to be
very energy-efficient --- academic and industrial investments in
neuromorphic computing have been steeply rising in the last
decade. Together with quantum computing, at this moment neuromorphic
computing is certainly the most intensely followed route toward
unconventional computing.

The field is diverse and can be segmented in several ways, for instance
\begin{itemize}
\item according to the targeted hardware platforms: digital
  simulations of neural dynamics like in IBM's by now almost classical
  TrueNorth microchip; or analog VLSI processors that instantiate
  neural spikes as physical electric pulses; or demonstrators of
  elementary neuro-dynamical phenomena in novel devices and materials
  \cite{Markovicetal20};
\item or according to how closely the computational models adhere to
  the biological original, which can range from adopting detailed
  neuroscientific models of neurons, circuits and adaptive dynamics to
  the extreme abstraction of calling an optical delay ring system
  ``neural'' when it used as a reservoir computing medium;
\item or according to the envisioned tasks which range from adaptive
  sensing, to nonlinear signal processing and control, to standard
  machine learning use-cases to complete cognitive agents;
\item or according to the computational architectures and algorithms,
  which besides all sorts of artificial neural networks and their
  learning procedures includes other ``cognitive'' models that are
  expressed in non-neural formalisms of machine learning and AI.
\end{itemize}

In this report we focus on an aspect that runs through all of these
directions of research (except maybe in fundamental materials and
device research), namely that most tasks require to integrate and
combine input information over several timescales which may be long
compared to the physical time constants afforded by the used physical
substrate.

\subsubsection{Unconventional computing}

The field of ``unconventional computing'' is even more diverse than
neuromorphic computing. This is already apparent from the fact that
over the decades work this field has been branded under many other
names like ``natural'', ``emergent'', ``physical'', ``in-materio'' computing
(we could list about twenty more) in different scientific communities
and with a variety of research objectives. In this report we use the
term ``unconventional computing'' in a broad way to include all such
approaches. We cannot attempt a systematic overview here and refer to
an introduction: \citeA{Stepney17}, a survey:
\citeA{EuropeanCommission09}, an extensive collection:
\citeA{Adamatzky17ab}, and an attempt at a theoretical unification:
\citeA{Jaeger21a}).

Most unconventional computing approaches aim at finding non-symbolic
concepts of ``computing'' that open alternatives to Turing-equivalent
algorithms which can be realized in non-digital physical substrates. A
key methodological principle to think about an unconventional
computing system is \emph{``to exploit the physics of its material
  directly for realizing its operations''} \cite{Zauner05}, which means
that the information which is being processed becomes instantiated in
physical quantities of any useful kind (electronic, magnetic,
chemical, biological...), and physical dynamics of any kind (energy
dissipation, wave propagation, pattern formation, phase
transitions...) are 1-1 interpreted as computational operations.

Such enterprises are distinguished from the niche of theoretical
computer science and physics called ``hypercomputation''
\cite{Ord06}. Unconventional computing aims at novel computing
concepts and technologies that are \emph{alternatives} to digital
computing in that the very definition of what ``computing'' could be is
changed. In contrast, hypercomputation seeks for ways that enable
\emph{extensions} of Turing computability in that symbolic functions
become computable which no Turing machine can compute. Here the basic
definition of what a ``computation'' is --- namely, effectively
evaluating a symbolically defined function --- remains
unchanged. Although hypercomputing is sometimes subsumed under
``unconventional'' computing in the literature, we exclude
hypercomputation from our understanding of ``unconventional'' computing.

Neuromorphic computing is frequently subsumed under ``unconventional''
computing, too. We pay respect to the singular standing of
neuromorphic computing and treat it here as a separate field, because
today is by far more intensely researched and far closer to practical
exploits than all other unconventional computing approaches taken
together. We observe that due to its promises and successes,
``neuromorphic'' has become a terminological attractor: we sometimes
observe that foundational research in unconventional materials and
nanoscale dynamical phenomena tends to be called ``neuromorphic'' when
the linking with neurons and brains seems rather far-fetched.

\vspace{0.5cm}

All three fields that we consider --- deeply scaled digital,
neuromorphic and unconventional --- face challenges related to
timescales which are not present in classical digital
computing. These challenges present themselves in many different ways
with regards to response latencies, information throughput
constraints, achievable memory spans, variable speeds of processing,
delays, accuracy versus time budget tradeoffs, hardware possibilities
versus task requirements, aging, and many more --- not to forget (as
we will see) the challenge of quantifying ``time'' in the first
place. In this long article we attempt to give a transparent navigation
guide through this bewildering landscape.

\subsection{ ``Computational extension of timescales'' can mean many
  things}

The first step is to become aware that ``computational extension of
timescales'' is not a clear-cut single problem. There are
numerous scenarios and reasons why or how one would want to manipulate
timescales in some way or other. Some examples:

\begin{itemize}
\item One may wish to use an analog electronics processor, whose
  physical time constants are very small, in a ``slow'' online signal
  processing task which requires long memorizing and integration times
  of incoming information. This was one of the starting motivations for
  the MemScales project.
    
\item In online signal processing tasks, one may need to adapt the
  ``speed'' of the processing system to faster or slower sorts of
  input signals.
  
\item In a speech recognition task, one needs a hierarchy of
  short-term and  working
  memory timescales that range from the millisecond range of phonemes
  to syllables to words to phrases to sentences to textual contexts
  that in turn range from a few sentences to lines of argument to an
  entire narrative. Furthermore, the processing of most of these
  intermediate timescales  needs to access information that is
  stored in some long-term memory. Similar multi-timescale support is required for many signal processing tasks, including imaging, audio, video, graphics, sonar, radar etc.

\item While memory timescales reach into the past, many cognitive
  processing tasks likewise require to reach out into the future
  across several timescales. For example, in control architectures for
  autonomous mobile robots one needs anticipations of the effect of
  current actions on the short timescale of avoiding obstacle
  collision in arm motion generation, as well as on several long
  timescales of purposeful action planning. 

\item Incoming sensor information or user input can arrive at varying
  levels of ``information density per time unit''. In a realtime
  computer game, the human gamer may at some times do nothing for 
  seconds or minutes, while at other times the gamer may fusillate the
  game engine with  volleys of joystick commands. In those periods of
  high input density the engine must in some way process the incoming
  information faster than when the user is idling. Similar high
  variations in computational load occur in many other applications,
  for instance in fault monitoring or online decision support
  systems, and for a broad range of body-area-network oriented sensors
  which are vital for  future healthcare applications.

\item So-called \emph{anytime} algorithms in classical digital
  computing are designed to output possibly inaccurate results fast
  when immediately needed, but would continue processing and generate
  more accurate results when given more time. ``Cognitive'' processing
  procedures in neuromorphic and unconventional systems might perform
  similarly by running several processes in parallel and in different
  subsystems, fast ones for approximate results and slow ones for more
  deeply ``thought-out'' results. This may be related to the common
  idea in cognitive architecture research that fast responses are
  obtained from a first ``feedforward'' bottom-up pass through a
  neural processing hierarchy, while on a longer timescale top-down
  pathways become activated which lead to recurrent ``deliberations''
  about an appropriate system response (as for instance in adaptive
  resonance theory \cite{Grossberg76, Grossberg76a}).
\item Related to the previous two points:
  online processing of input signals may be sensitive to the current
  information density in the input and workload of internal
  processing, even at the level of individual signal values. In
  assuring a ``correct'' system behavior, pure worst-case behaviour
  analysis is likely inappropriate in neuromorphic, unconventional and
  deeply scaled systems, among other
  due to the already mentioned strong variability of the underlying
  hardware. The ratio of achievable versus desirable levels of
  ``correctness'' will be time-varying, constituting a timescale
  dynamics that is important for qualifying the performance of a
  computing system.
\end{itemize}

This ad hoc list of scenarios illustrates that there is not a single
generic problem of ``timescale extension'', but a whole spectrum of
such problems.  The specific requirements are moreover application-
and context-dependent, which makes it even harder to derive a 
theoretical basis spanning this broad domain.

\subsection{When does a physical system compute?}\label{secWhatIsComputing}

In order to get a clean and practically useful understanding of what
``computational extensions of timescales'' can actually mean, we first
have to understand what we mean when we say that a physical system
``computes''. In this subsection we thus discuss the question ``when
does a physical system compute?''

This question is in fact the title of a foundational paper by
\citeA{Horsmanetal14}. The authors analyze the conditions when and in
what sense one can claim for a physical system that it
``computes''. Their main conclusion is that ``computation''
is defined in an abstract formal-mathematical sphere (for instance
framing computing as sequence of Boolean operations or as the update
rules of a Turing machine). Inputs and outputs for computations are
defined in this abstract sphere (for instance as a sequence of 0's and
1's in digital computing, or as real-valued timeseries in analog
signal processing). In order to link this abstract casting of
``computing'' to an underlying physical machine, the abstract, formal
inputs $u$ and outputs $y$ must be mapped to the physical signals that
enter and leave the machine.

Horsman et al.\ call this mapping the \emph{representation} relation
between the physical and the abstract domains. Importantly, the
representation relation is neither established by the physical
machine, nor is it part of the abstract model of computing. Instead,
the representation is effected by and incorporated in the external user of the
system. This user calls upon practical judgement based on community
consensus to warrant that the physical I/O machinery (like keyboards,
screens, oscilloscopes) realizes the representation mapping in an
acceptable way. The external user in which the representation becomes
incorporated is almost always a human, but it could in principle also
be another intelligent agent (animal or even a future
machine). Horsman et al.\ call such agents  \emph{computational
  entities} and define them to be \emph{the physical entities that locate
  the representation relation}. We will use the more common term
``user''.

It happens within the user that, when a computer prints ``{\sf the solution
is x = 2}'' on its screen, this physical signal becomes \emph{decoded}
to the abstract output format (here the integer 2). This decoding
involves epistemological conditions which philosophers find hard to
sort out, including
\begin{itemize}
\item the neural instantiation of the abstract model (here: the
  integers) in the user's brain (or is it the cognitive instantiation
  in the user's ``mind''? don't start asking philosophers!),
\item and social processes of communication among all potential users
  of the system to reach a consensus that this visual reading of
  screen outputs is an acceptable and universally shareable decoding
  procedure.
\end{itemize}
While we all have become used to the various ways that inputs and
outputs are given to and taken from digital machines, to a degree that
none of us will question these procedures, the situation gets less
clear when it comes to unconventional new kinds of computing
machines. For example, when a slime mold in a petri dish grows
branches that extend into higher-concentration areas of a nutrient (a
popular model of unconventional computing with a substantial
literature \cite{Adamatzky18}), how can one read out an abstract
output like a Boolean ``True'' from this physical system?

\begin{figure}[htb] 
\centering
\includegraphics[width=14cm]{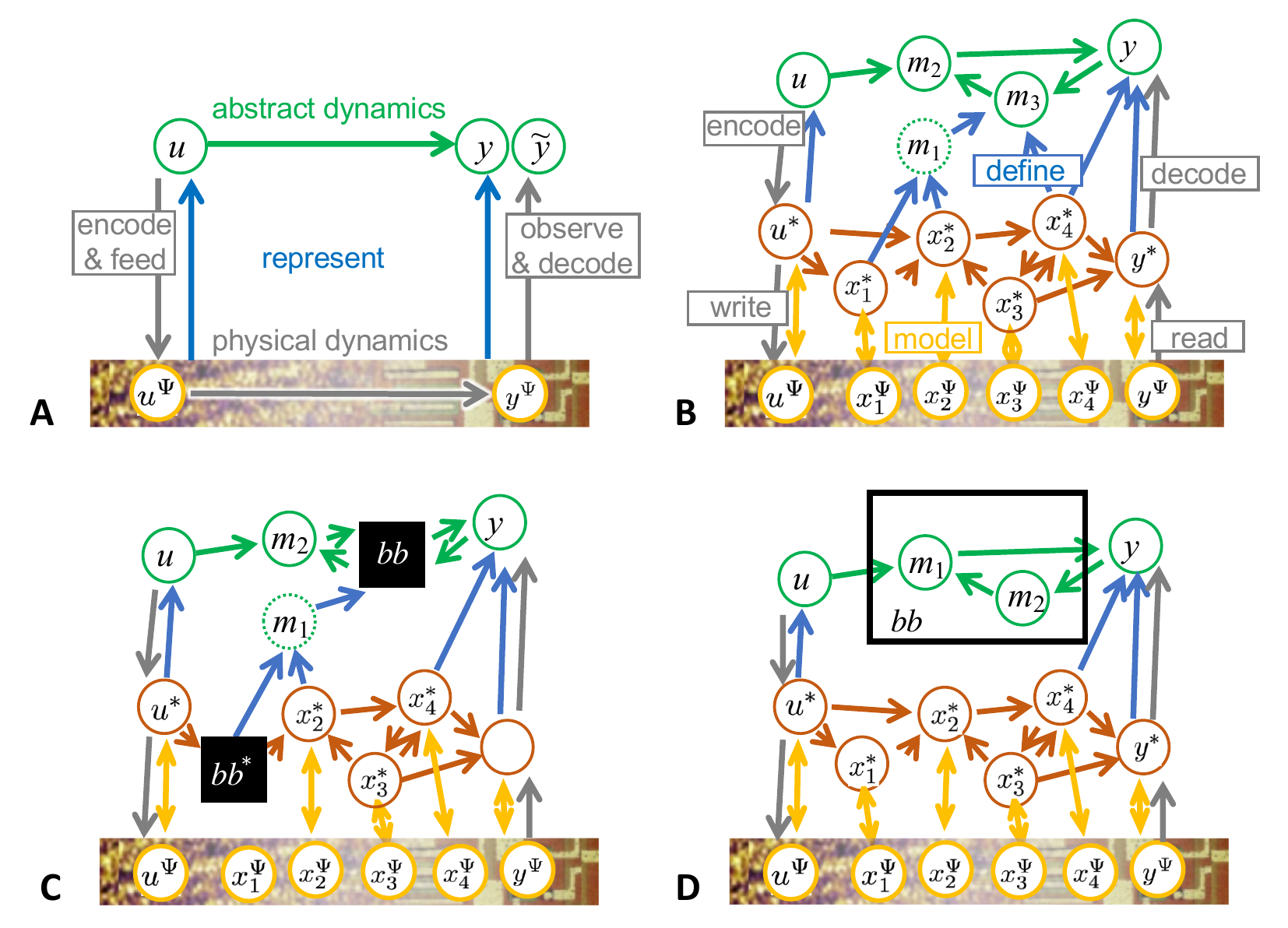}
\caption{Increasingly complete accounts of a computing
  system. \textbf{A:} The elementary view of
  \protect\citeA{Horsmanetal14}. \textbf{B:} An idealized schema of a more
  complete modeling structure, comprising an analytical system model
  which is a reflection of the physical system, on top of which an
  abstract model of the input-to-output transformation is defined by
  variables that represent possibly complex subprocesses in the
  analytical model. \textbf{C:} Gaps in the analytical and abstract
  models can be filled with experimentally obtained blackbox
  models. \textbf{D:} The entire abstract input-to-output
  transformation can be modeled with a hand-crafted blackbox model.
  For explanation see text.}
 \label{IOcompleteVsNot}
\end{figure}

\citeA{Horsmanetal14} complete their account by assuming that the
abstract domain includes some model (not further specified by the
authors) of an abstract dynamical process which transforms the input
into the output, which together with the physical dynamics leads to a
diagram with four arrows which should commute (Figure
\ref{IOcompleteVsNot}\textbf{A}). That is, when the abstract dynamical
model transforms input $u$ to output $y$ (green arrow in the figure),
and when the abstract input $u$ is first \emph{encoded} in a physical
input signal $u^\Psi$, then physically transformed to a physical
output signal $y^\Psi$, which is finally \emph{decoded} to an abstract
output $\tilde{y}$ (grey arrows in Figure
\ref{IOcompleteVsNot}\textbf{A}), the abstractly determined output $y$
should be (approximately) equal to the output $\tilde{y}$ obtained via
the route through the physical machine.

We emphasize that according to this view of what makes a system
``compute'', a computing system necessarily must have some provision
to accept input and produce some output. This commitment seems natural
enough, but lies in opposition to certain foundational proposals in
theoretical physics which attempt to reduce all physical laws to
elementary information processing operations --- the ``universe as
computer'' view \cite{Zuse82, Zuse93, Wheeler89, Lloyd13, Zenil13,
  DeutschMarletto15}.  In contrast, we follow \citeA{Horsmanetal14}
and other theorists in that a closed physical system cannot be said to
``compute''. 

\subsubsection{Analytical system model}

While we use the approach of \citeA{Horsmanetal14} as a starting
point, we find that the account given in that work lacks too much
detail for our purposes. In engineering and modeling practice, the
picture becomes enriched with further elements (Figure
\ref{IOcompleteVsNot}\textbf{B}). The most important addition is an
intermediate \emph{analytical system model} inserted between the
abstract model and the physical machine, turning Horsman et al's
two-layer model into a three-layer one. This intermediate model
formally reflects the physical processes which transform the physical
input $u^\Psi$ into the physical output $y^\Psi$. The input,
intermediate state, and output variables $u^\ast, x_i^\ast, y^\ast$ of
the analytical physical model correspond to measurable physical
quantities $u^\Psi, x_i^\Psi, y^\Psi$. The relation between the
physical quantities $v^\Psi$ (where $v$ means $u$, $x$ or $y$) and
their analytical correlates $v^\ast$ is again effected by the human
engineer and involves physical measurements, knowledge of physical
laws, and community consensus about adequacy criteria (for instance
which measurement apparatuses and procedures are admissible according
to which accuracy demands). The relationship between the physical
hardware system and the analytical model is the \emph{modeling}
relation from the empirical natural scienes. According to the 
scientific criteria of modeling reality in the natural sciences, as
decreed by \citeA{Popper59}, the analytical model should enable
nontrivial predictions which can be empirically checked by physical
measurements. 

An analytical system model usually reflects its physical target system
bidirectionally in the following sense: each model variable $v^\ast$
corresponds to a single physical state variable $v^\Psi$, which is
(in principle) measurable;
the model lets its variables evolve with regards to a continuous time
variable $v^\ast (t)$ which is measured in standard physical units
like seconds; model variable trajectories $(v^\ast (t))_{t \in T}$ can
be matched against physical state trajectories
$(v^\Psi (t))_{t \in T}$ by comparing $v^\ast (t)$ with physical
measurements $\mathcal{M}(v^\Psi (t))$ timepoint by timepoint (and
where time itself is measured in the physical world by a physical
clock). Model trajectories $(v^\ast (t))_{t \in T}$ can \emph{predict}
physical trajectories $(v^\Psi (t))_{t \in T}$. Such predictions about
measurable reality are a critical property of system models in the
natural sciences.

The most common mathematical format for analytical system models are
ordinary differential equations (ODEs). This is the mathematical
language which is typically used in electronic circuit engineering. It
is also a formalism of first choice in many models in computational
and theoretical neuroscience. In materials science one also often uses
partial differential equations (PDEs) or finite elements
formalisms. Furthermore, stochastic versions of all of these may also
be employed. These are continuous-time formalisms. Discrete-time
formalisms (like iterated maps) or discrete-value formalisms like
finite-state Markov models or Petri nets, or discrete-time,
discrete-space Markov random fields, are also in use, though more
rarely in our perception.  However, in this report we will generally
base our discussion on ODE analytical models, which are continuous in
time and value.

When we speak of an analytical model, we refer to its
mathematical formalisation. All of the above can be approximately
simulated on digital computers, which entails discretization in time,
space and value. Those digital simulation formalisations (and the
problem of guaranteeing approximation bounds) are obviously important
in practice, but they are not what we mean by analytical models.

The analytical system model serves a number of important functions:

\begin{itemize}
\item It is the interface to the physical machine which is used by the
  developer of abstract computational procedures. Abstract computing
  procedures are rarely (if ever) matched directly against the
  hardware --- programmers do not normally use oscilloscopes to test
  or debug their programs.
\item It is used by the hardware engineer to design and analyse
  physical systems.
\item It can be used to simulate the physical system on a digital
  computer. This has become an indispensable routine for hardware
  developers, and in unconventional computing simulations will
  likewise be
  needed by designers of abstract ``algorithms''.
\end{itemize}

Like any physical system model, analytical system models abstract from
the physical reality to some extent. For instance, in the practice of
electronic microchip engineering, the analytical system model that is
used by the chip designer may be expressed in sampled-signal form
with implicit assumptions on clock synchronization. The defining
characteristic of analytical system model is not that it captures the
``real'' physical system in every detail, which  is
impossible. Rather, it is defined by its \emph{purpose}, namely that
it should model the actual physical behavior of the target system at
the necessary level of accuracy that is needed for understanding its
computational exploits.

\subsubsection{Abstract computational model}

The abstract computational model sits on top of the analytical system
model. It explains an abstract computation $u \to y$ through a process
between intermediate abstract state variables $m_j$. These abstract
state variables are \emph{defined} by formal
derivation from analytical model
variables $v^\ast$, possibly in cascades of intermediate meta
variables, like $m_1$ in Figure \ref{IOcompleteVsNot}\textbf{B}. In
various contexts, such meta variables are called by names like
``features'', ``abstractions'', ``characteristics'', ``descriptors'',
 ``transforms'' or the like.

 This picture (as in Figure \ref{IOcompleteVsNot}\textbf{B}) is still
 a simplification. In real-world practice one finds significant
 variations of this idealized scheme, which will become relevant when
 we later discuss computational timescale extensions:

\begin{itemize}
\item The abstract modeling layer often is further sub-structured in
  layers of increasingly abstract models. A familiar case are
  compilation hierarchies in digital programming, where at the lowest
  sub-layers one finds microprocessor instruction sets, then assembler
  code at the next higher level of abstraction, followed by further
  layers that are expressed in increasingly ``higher'' programming
  languages, until at the top one may find graphical user
  interfaces. For computing with (possibly analog) spiking
  neuromorphic microprocessors, such abstraction hierarchies are
  beginning to be developed \cite{Zhangetal20}.
\item When reservoir computing (\emph{echo state networks}
  \cite{Jaeger01a} and \emph{liquid state machines}
  \cite{Maassetal01}) methods are used to compute with unconventional
  physical substrates, and in other fields of unconventional
  computing, an analytical physical model is often not available. The
  abstract computational model then includes one or several ``black
  boxes'' which are matched to the underlying hardware not via formal
  definition from an analytical physical model but via model
  estimation methods of machine learning, statistics or signal
  processing.

  Besides fixing blackbox models purely by machine learning methods
  based on experimental observations, they may have some hand-designed
  internal structure that exploits prior hypotheses about causal
  mechanisms in the physical system, as illustrated in Figure
  \ref{IOcompleteVsNot}\textbf{D}. Pre-configuring trainable (sub)system
  models with such \emph{structural bias} may greatly improve the
  trainability of these models from measured data.
\item Also the analytical system model may have gaps
  that are filled with experimentally determined blackbox
  models (Figure \ref{IOcompleteVsNot}\textbf{C}).

\item While Horsman et al.\ consider a single user who embodies the
  representation relation, in real life this embodiment will often be
  split over several human experts each of whom is taking care of only
  one step in an abstraction hierarchy as in Figure
  \ref{IOcompleteVsNot}\textbf{B} --- for instance, an electronics
  engineer takes care of representing a physical machine by an
  analytical model; a microchip architecture designer creates the next
  abstraction level with a machine instruction set; and so forth
  upwards in the abstraction ladder up to a web designer creating a
  webstore interface. All these experts must communicate with at least
  the next expert below and the expert above. The further such
  mutual understanding reaches out across levels, the more seamless
  become the operations of the overall multi-agent ``computational
  entity''. In the digital computing world, such distributed but
  mutually informed division of expertise has become well established
  over the decades that this field could evolve. Progress in
  unconventional computing technologies is still hampered by
  disciplinary boundaries between level specialists, like between
  materials scientists and machine learning experts.
\item In theoretical computer science there
  are two kinds of formal models of computing systems. The first kind
  is the one that we here called abstract computational models. These
  models include computer programs expressed in programming languages
  on all levels of the compilation hierarchy and machine- and
  compiler-independent abstract formalizations like Turing machines or
  random access machines. They describe effective
  \emph{mechanisms} or \emph{procedures}, and they can be effectively
  translated and compiled down to machine instruction sets that can be
  executed on real machines. The second kind of formal models in
  symbolic computing theory do not describe the ``mechanics'' of
  computing processes but their \emph{semantics}. These models are
  expressed in mathematical logic formalisms and relate the data
  structures and transformations of the abstract computational model
  to external task specifications. In \citeA{Jaeger21a} we have called
  these two kinds of models \emph{how-} and \emph{what-}models, and have
  provided a more in-depth analysis of the two. In this report we do
  not further consider the second kind of models.
\item The distinctions between the physical machine, the analytical
  model and abstract meta models are not as clear-cut as we suggested
  here. Consider, for example, an FPGA processor. When it is
  programmed (better word: configured) in two different ways for
  use in two different applications, its Boolean circuitry is in a
  sense ``hardwired'' in two different ways. Would we say that these
  are two different physical machines, needing two different
  analytical models? Or more generally, if any digital machine
  includes non-volatile memory devices, like a harddrive or a BIOS
  memory, and these devices are re-set, one has permanently (though
  usually reversibly) changed the physics of the machine. Is a new
  analytical model needed? At this moment we see two principled
  options to reconcile such scenarios with our accounts from Figure
  \ref{IOcompleteVsNot}:
  \begin{enumerate}
  \item Option 1: Observing that in digital machines the physically
    changeable memory elements together with their local control
    circuits are essentially bistable, a single analytical model which
    captures the bistable attractor dynamics of the memory elements
    will cover all the possible non-volatile configurations.
  \item Option 2: Non-digital machines may be subject to persisting
    physical changes which cannot be cast as attractor dynamics. For
    instance, the graded resistance states of filamentary or
    phase-change memristors would rather be mathematically described
    as transients which are so slow that for practical purposes they
    become constant.  Such non-attractor, very slow physical
    changes could either become modeled in a highly resolving
    analytical model, as in option 1. Or, a quite different option would
    be to capitalize on the reversibility of such physical changes and
    mathematically cast a physical system as an equivalence class of
    reversible configurations. This may be more
    mathematically insightful than option 1.
  \end{enumerate}
\end{itemize}

The correspondence between physical state variables $v^\Psi (t)$ and
their formal reflections $v^\ast (t)$ is bijective in the sense
mentioned in the previous subsection.  The correspondence between the
abstract input, meta and output variables $u, m, y$ and variables
$v^\ast$ in the analytical model is of a different kind. Firstly, a
single abstract variable can be formally derived from several
analytical model variables (like $m_1$ and $y$ in Figure
\ref{IOcompleteVsNot}), and a single analytical variable can enter the
derivation of several abstract variables (like $x_4^\ast$ in the
figure).  Secondly, abstract model variables may temporally evolve in
other ways than by following physical time $t$ --- this will be
discussed in the following Section \ref{subsecWhatTime}. As we will
see, having abstract ``kinds of time'' that differ from physical time
$t$ is a key for computational extensions of timescales.

An abstract computational model need not and usually cannot predict
the analytical system model from which it is derived. The abstract
model will often only exploit only a fraction of the physical
phenomena that are captured by the analytical model, and the abstract
model may contain subprocesses that have no apparent counterpart in
the analytical model. The word ``model'' in ``abstract computational
model'' does not refer to modeling the physical system (via the
analytical model) in the sense of natural science modeling. When we
speak of a formal Turing machine as a ``model'', we suggest that it is
one possible way (one ``model'' among many) to formalize a specific
input-to-output transformation. Confusion is added by the fact that
sometimes, however, we \emph{do} speak of an abstract computational
model as a model of a physical system, as in the case of the random
access machine model, which is tailored to match a physical
von-Neumann machine in a halfway realistic manner. But in general, the
relation between an analytical physical model and an abstract
computational model is unidirectional and partial: some of the
abstract meta variables are formally derived from some of the
analytical variables. As we will discuss in
subsection~\ref{subsecTrafoCompl}, not every abstract computational
model that solves an externally specified task will be realizable on
the basis of a given hardware system. Given an externally specified
task and a fixed hardware base, there may or may not be some abstract
computational model that solves the task \emph{and} can be defined
from the analytical system model.

There is one limit to the freedom of decoupling an abstract
computational model from an analytical system model: the abstract
input $u$ must be encodable in analytical input variables and
definable from them, and the
abstract output $y$ must be decodable from analytical output
variables. 

From an analytical system model one can define an unlimited multitude
of different abstract computational models. Some of them can be
related to each other in abstraction hierarchies (for instance,
compilation hierarchies in digital programs). But they can also
represent incomparably different approaches to do useful
``computations'' on a given hardware basis with its analytical system
model. For instance, a piece of hardware whose analytical model can be
understood as a recurrent spiking neural network with adaptable
synaptic connections, could be used as a basis for abstract
computational models of
\begin{itemize}
\item \emph{reservoir computing:} adapt only the readout synaptic
  connections by some algorithmic procedure that leads to a linear
  regression \cite{Heetal19}. 
\item \emph{unsupervised STDP learning} which combined with a
  maximum-activity detection operation on selected neurons yields a
  classifier \cite{Yousefzadehetal18, Covietal18},
\item \emph{gradient-descent learning} using a spike-adapted version
  of backpropagation through time \cite{YinCorradiBohte21},
    
\end{itemize}
just to list some examples of work done in the MemScales consortium.

\subsubsection{Transformation-completeness and model
  viability \label{subsecTrafoCompl}}

We call an abstract model
\emph{transformation-complete} when it mathematically fully specifies
how abstract outputs $y$ result from inputs $u$. Abstract
computational models used in computer science and neuromorphic
computing are typically transformation-complete (while computational
models in the neurosciences often are not because they leave some
transformations under-specified). Transformation-complete
models can be simulated on a digital computer.
  
A transformation-complete model can include blackbox models. The meta
variables inside blackbox components are not derived by definition
from analytical model variables. In an extreme case the entire
$u \to y$ transformation is contained in a single blackbox model
(Figure \ref{IOcompleteVsNot}\textbf{D}). 
  
  \begin{figure}[htb] 
\centering
\includegraphics[width=6cm]{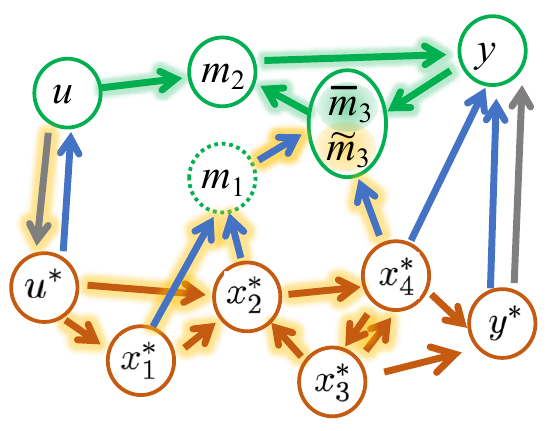}
\caption{Twofold determination of meta-variables $m$ through
  dependency pathways within the abstract computational model (green
  shadows) and via the enconding, analytical model transformations,
  and the definition operations (orange shadows). Figure shows this
  double determination for $m_3$.}
 \label{figCommute2}
\end{figure}
  
A transformation-complete abstract computational model must be
\emph{valid} in the following sense. First notice that the value of
each meta variable $m$ is determined in two ways: firstly, by its
ultimate dependence on the abstract input $u$, which is mediated
through all paths within the abstract model that lead from $u$ to
$m$. Let us call the version of $m$ which is determined in this way by
$\bar{m}$. In Figure \ref{figCommute2}, which is an excerpt from
Figure \ref{IOcompleteVsNot}\textbf{B}, we pick $m_3$ to illustrate
this. Here $\bar{m}_3$ depends on $u$ via the paths underlaid with
green shadows. Secondly, $m$ is also determined by its formal
derivation from variables in the analytical system model, which in
turn are also dependent on the abstract input signal $u$ via the input
encoding operation from $u$ to $u^\ast$ followed by all pathways
inside the analytical model which lead from $u^\ast$ to variables
$x_i^\ast$ which then are used to derive $m$ (these pathways have
orange shadows in Figure \ref{figCommute2}).  Let us call the version
of $m$ which is determined along these second paths as $\tilde{m}$.
  
The two versions of $m$ must approximately be the same,
$\bar{m} \approx \tilde{m}$. It is a delicate issue how
``approximately'' should be defined. It will require a case by case
agreement about what tolerance is admissible such that the computations
defined by a transformation-complete abstract computational model in
terms of variables $\bar{m}$ are ``well enough'' served by the replay in
variables $\tilde{m}$ which are derived from the analytical model. The
analysis becomes particularly difficult when there are recurrent
cycles in the dependency paths (like the cycles
$m_2 \to y \to m_3 \to m_2$ or $x_3^\ast \to x_4^\ast \to x_3^\ast$ in
the figure). Such recurrent dependencies incur the danger that small
value mismatches between $\bar{m}$ and $\tilde{m}$ can become
magnified over time.

In the world of digital computing, a ``good enough'' match between
abstract bit variables $\bar{m}$ with values 0 or 1, which are
defined from analytically modeled gate voltages
$x_i^\ast [\mbox{\sf Volt}]$ as $\tilde{m}$, is ensured by three
conditions:
  \begin{itemize}
  \item The definition operation
    $x_i^\ast [\mbox{\sf Volt}] \to \tilde{m}$ is a binary
    thresholding operation which is insensitive to real-valued minor
    variations of $x_i^\ast [\mbox{\sf Volt}] \to \tilde{m}$.
  \item The recurrent dynamics of an electronic Boolean circuit model
    is a bistable attractor dynamics which makes
    $x_i^\ast [\mbox{\sf Volt}] \to \tilde{m}$ converge fast
    toward two widely separated fixed points.
      \item The global clock within the analytical model gives
        temporal reference points (in the middle of a clock cycle, for
        instance) when the definition
        $x_i^\ast [\mbox{\sf Volt}] \to \tilde{m}$ is to be carried
        out, and the electronic bistable dynamics will be safely
        converged closely enough to either of the two Volt levels such
        that the thresholding operation will yield the correct 0 or 1
        value.
  \end{itemize}
  
  We call the combination of an analytical physical model and a
  transformation-complete abstract computational model \emph{valid} if
  the values of meta variables $\bar{m}$, as determined by processing
  input within the abstract model, are ``well enough'' approximated by
  the defined $\tilde{m}$. This means that the diagrams in Figure
  \ref{IOcompleteVsNot} must approximately commute. This is the
  essential message from \citeA{StepneyRasmussenAmos18}. The devil is
  in the detail: these diagrams will rarely commute in mathematical
  exactness, and which degrees (and which sorts) of mismatches between
  $\bar{m}$'s and $\tilde{m}$'s make the abstract model function as
  desired on the basis of the analytical model needs to be studied
  anew for each case as long as we do not have a general theory which
  quantifies approximaton errors and relates these errors to formal
  task models.

  A further challenge is that the analytical model must capture the
  real physical dynamics ``well enough'', too. The definition operations
  here become experimental measuring operations. We do not pursue this
  issue in more depth here.

  \subsubsection{From task to abstract computational
    model}\label{subsecTask2Model}

Here we briefly insert some remarks on the relation between externally
specified tasks, or ``use cases'', and abstract computational
models. While this topic is outside the scope of this report, it is of
great practical importance.

A computational task may at first be specified only in rather vague
natural language by some end-user or customer. In order to let it be
solved by a computing system, the first step is to distil from this
initial, informal specification a mathematically rigorous one. This is
a nontrivial process, for which an array of systematic methods has
been proposed in the digital software engineering world, invoking for
instance methods from knowledge acquisition, UML models or user
interface designs. Depending on how one goes about this initial
formalization stage, one obtains a processing model formalized in some
formalism, for instance a dataflow diagram.

This formal
task model is not yet an abstract computational model, in that it
usually does not relate to procedures by which the task could be
``computed''. Thus in a second step one must design a
transformation-complete abstract computational model which, on the one
hand, solves the task and which, on the other hand, can be defined
from the analytical model of the hardware system that one wishes to
employ.

In the digital world this second step usually ends in the writing a
computer program after possibly some systematic intermediate task
representations and thinking about which programming paradigm (like
imperative versus declarative) and programming language is most
appropriate --- all of that is taught to computer science students in
software engineering courses. The resulting program is an abstract
computational model.

In the practice of digital computing, this second step is also the
last one. The rest --- compiling the high-level program down to
machine code through a sequence of increasingly hardware-reflecting
abstract models and ultimately ``running'' the program by creating
physical input signals --- is a well-established technology that is
hidden from  end-users who buy it together with the computer
and its operating system. The more we start thinking about how easy
this is for users and high-level programmers, the more we should feel
amazed. This is the magic of digital computing: general-purpose
digital computers come with pre-installed abstract computational
models (the operating system) that are equivalent to a universal
Turing machine and therefore can execute any program written in any
programming language. The challenges for the user are not principal
problems of feasibility versus impossibility, but of optimization:
clean coding, efficiency, robustness, maintainability and so forth.

The situation is far more difficult in neuromorphic or other
unconventional computing systems. We have not yet discovered a formal
model of a ``universal'' computing in such systems. Every physical
system is unique and could ultimately realize (only) a specific family
of tasks.  It becomes a principal problem to find out whether a formal
task specification can indeed be translated at all to an abstract
computational model that is supported by a given hardware basis.

Similar difficulties also arise with deeply scaled digital systems,
though not for end-users (who will still buy machines together with a
Turing-equivalent operating system), but for the operating system
designers. As we mentioned in Section \ref{subsecDeeplyScaled}, the
further miniaturization is pushed, the more the hardware exhibits
problematic static and dynamical variability effects that shatter the
clean picture of a reliable symbol processing engine, and that need to
be identified, monitored, controlled and compensated in an elementary
abstract computational layer that lies between the analytical model
and the operating system. We called it the nanoprogramming layer for
lack of a better word. 

However, also in the digital domain the step from a formal task
specification to a computational model can be problematic. Sometimes
(especially in scientific computing and signal processing and control
using DSP microchips) the formal task specification comes in the form
of a continuous-time, continuous-value, mathematically
infinite-precision model like a set of ODEs. Because digital computers
do not admit infinite-precision numerics, one has to solve the
problems of accuracy and stability that come with data quantisation
\cite{Menardetal10,Menardetal19}.

\subsection{What time is it, and where?}\label{subsecWhatTime}

So far we have developed a three-level picture of computing systems,
with the physical system at the basis, abstract computational models
at the top, and an analytical physical system model in between. For
the theme of this article is it important to recognize that ``time''
may be conceptualized, quantified  and formalized differently on these
levels.

The physical system evolves in physical time -- that continuous arrow
of change that physicists and ordinary people take for granted
and that philosophers have not an easy time with \cite{Callender11}.

Analytical system models use the symbol $t$ to capture the continuous
physical time in their ODEs, and the dot in $\dot{\mathbf{x}}$ means a
rate of change that is quantified with respect to real
seconds. The formal system trajectories
$(\mathbf{x}(t))_{t_{\mbox{\tiny min}} \leq t \leq t_{\mbox{\tiny
      max}}}$ can be aligned to physical reality by human experts with
physical measurements and stopwatches, and there seems little to
debate. 

However, when it comes to the abstract computational models, there are
many ways for time to enter the game.

We start with a startling observation: in the textbooks of theoretical
computer science one will not find the word ``second''. The update
steps of a Turing machine, or the sequence of algorithmic
transformations of symbolic data structures, are not connected to
physical time. This has a deep reason: the theory of symbolic
computing has historically emerged from the study of logical
inference. Turing invented the Turing machine not as a model of a
physical machine, but as a model of the logical reasoning steps that
lead a mathematical reasoner from one argument in a proof to the
next. One can say that the forward progression in the execution of
symbolic algorithms steps from Truth to Truth, --- not from Timepoint to
Timepoint. The decoupling from physical time is even explicitly stated by
\citeA{Turing36}: \emph{``It is always possible for the computer to break
  off from his work, to go away and forget all about it, and later to
  come back and go on with it''} --- provided that ``the computer''
(a human male in Turing's famous article) left a written note to
remind him later at which stadium in the computation he took a
break. In modern digital-technology parlance: after writing a bit into a
memory cell, the bit stays there for arbitrary (physical-real)
timespans until it is read again or overwritten. The availability of
non-volatile discrete state changes in digital computers allows
digital computations to decouple from physical time.

On the other hand, there are abstract computational models that
\emph{do} use the ``$t$'' of physicists. In particular we think of the
signal transformation procedures which were designed in the fields of
analog signal processing and control before the advent of digital
signal processing technologies, or in the field of analog
computing.

We will use the term \emph{mode of progression} to refer to the way
how the successive transformations of meta variables in the course of
a computational process can be seen as ``some kind of time''. We use
the fraktur letter $\mathfrak{t}$ to denote \emph{any} mode of progression.

In between the physically measurable time $t$ on the one end, and the
purely logical inference updates on the other end of a spectrum which
ranges from physical-temporal to logical-atemporal, there are abstract
computational models which use intermediate or hybrid modes of
progression, for instance
\begin{itemize}
\item sampled timesteps $n\,\Delta t$ in signal
  processing theory (still synchronized with physical time $t$);
  \item sampled timesteps as before, but augmented by a stochastic
    component (e.g. additive noise $n\,\Delta t + \nu_n$) to account
    for jitter;
\item dimensionless ``unit'' timesteps $n$ in iterated function models
  of computing, for example discrete-time recurrent neural networks;
\item models of information processing based on hybrid
  logical-physical modeling formalisms used in
  the formal verification of hardware-embedded digital computing
  systems \cite{Geuversetal10}, where logical argumentation steps are
  intertwined with analytical physical models of components of the
  complex hardware system that is being modeled;
\item models of real-time operating systems (RTOS) for digital hardware where
  certain computational processes (expressed algorithmically as a
  sequence of logical update steps) must not exceed a limited number
  of update steps due to latency constraints on the RTOS, such that this number multiplied with the (physical)
  clock period does not exceed a physical time limit;
\item Petri net models of concurrent information processing systems which
  come in many variations with regards to time models, from the
  classical purely discrete algorithmic version through timed variants to
  variants that involve continuous flows of tokens, defined with
  respect to continuous but arbitrary (not physical) time
  \cite{AllaDavid98};

\end{itemize}

In summary: while the physical computing system evolves in the
real-world physical time, and the analytical physical system model
must describe state evolutions with respect to the measurable $t$
(measured in seconds), ``anytime goes'' for the  modes
of progression of meta-variables in abstract computational models.

This view leads to an interesting issue which is of importance for the
theme of this report.  Given (i) that the analytical system model uses
physically measurable time $t$ for monitoring the evolution of its
analytical variables $v^\ast$, and (ii) that in the abstract
computational model other modes of progression may be used for the
meta variables $u, m, y$, the question arises how one can formally
change the mode of progression when one defines abstract meta
variables from analytical variables or other meta variables. There are
many ways. For illustration here is an ad hoc list of mathematical
operations to create new meta variables from existing analytical or
other meta variables. The first two definition methods in this list
are the two most widely used methods to obtain meta variables $m(t)$
that inherit the progression mode of physical time $t$ and thus are
\emph{synchronized} to the variables they are derived from:

\begin{itemize}
\item A meta variable $m(t)$ with the physical mode of progression $t$
  can be obtained through a \emph{function}
  $m(t) = G(v^\ast_1(t), \ldots, v^\ast_n(t)$ of analytical variables.
\item A meta variable $m(t)$ can be obtained through a \emph{filter}
  $(m(t))_{t \in \mathbb{R}} = H((\mathbf{v}^\ast(t))_{t
    \in\mathbb{R}})$ which transforms vector signals
  $(\mathbf{v}^\ast(t))_{t \in \mathbb{R}}$ to meta signals
  $(m(t))_{t \in \mathbb{R}}$.
\item Analytical model variables $v(t)$ or other meta variables
    $m(t)$, where $t$ is the physical time
    $t[\mbox{sec}] \in \mathbb{R}$, can be abstracted by dropping the
    commitment that the time parameter means the physical time that
    can be measured with physical clocks. To reveal this dramatic
    conceptual change in formal notation, we use $t^\emptyset$ as symbol for
    dimensionless continuous time. The abstraction then is formally effected by
    the simple replacement of $t$ by $t^\emptyset$. Abstract computational
    models that use dimensionless time can represent physical systems
    that run at arbitrary physical speeds.
\item By discretizing physical time $t$ through sampling with
    physical period length $\Delta [\mbox{sec}]$ one gets discrete
    time $n\Delta [\mbox{sec}]$ where $n \in \mathbb{Z}$. 
  \item When the hardware system is clocked but the clocking has
    physical time jitter, and when the analytical model correctly
    captures this jitter (deterministically or through a stochastic
    formalism), an abstract computational model that wishes to exploit
    the clocking can (i) opt for dimensionless clock increments $n$,
    putting the load of ``dejittering'' on the definition rule which
    must be robust to real-time clock cycle length variation; or (ii)
    it can use a clock increment model $\Delta_n [\mbox{sec}]$ with
    variable real-time increments where $n$ is the cycle counter
    index.
\item From discrete-time abstract models with progression modes
    $n\Delta [\mbox{sec}]$ or $n$ one can always define further
    discrete-time abstract models by subsampling, and sometimes by
    supersampling / interpolation. 
    
\item Discrete physical time $n\Delta [\mbox{sec}]$ can be
    abstracted to dimensionless discrete time $n$. Note that every
    analytical or abstract model which uses continuous time $t$ can be
    discretized by sampling, but the converse is not true: there are
    discrete-time system models (for instance, expressed in an
    iterated map formalism) for which no continuous-time corresponding
    system exists from which the discrete-time model is obtained
    through sampling. 
 
\item From analytical variable vectors $\mathbf{v}$ or meta variable
    vectors $\mathbf{m}$ which have modes of progression
    $\mathfrak{t} \in \{t [\mbox{sec}], t^\emptyset, n\Delta [\mbox{sec}],
    n\}$, one can define a new meta variable $m'$ by defining (i) a
    binary 0-1 \emph{trigger signal} $s(\mathfrak{t})$ which is
    zero most of the time and jumps to one whenever
    $\mathbf{v}(\mathfrak{t})$ or $\mathbf{m}(\mathfrak{t})$ in their
    evolution meet some trigger criterion, and (ii) a
    filter $M$ operating on the trajectories of
    $\mathbf{v}(\mathfrak{t})$ resp.\ $\mathbf{m}(\mathfrak{t})$. The
    new variable $m'$ has mode of progression $n$ and consists of the
    sequence of values returned by $M$ at times $\mathfrak{t}$ where
    the trigger signal is one. Alternatively, one may also keep a
    record of the inter-trigger intervals and include this information
    in the progression mode of $m'$, which would then take the format
    of a sequence of inter-trigger intervals of variable length. A
    familiar example is the abstraction of a continuous-time,
    continuous-valued neural membrane potential signal (as it may come
    out of the Hodgkin-Huxley equations for example) to a binary spike
    train signal. In digital real-time operating systems, the trigger
    signal can indicate the completion of a subprocess and $M$
    copies the result thereof. Event-based abstract
    computational models are generally defined from analytical
    models by such wait-trigger-read mechanisms.  
\item By various methods of memorizing/buffering, predicting, and
    time-warping one can locally stretch or compress a mode of
    progression $\mathfrak{t}$ such that a source progression
    $m(\mathfrak{t})$ becomes bidirectionally
    mapped on a target progression $m'(\mathfrak{t})$ in the same
    mode. There is a multitude of options which would need to be
    discussed in detail, depending on the specific formats of
    variables and their modes of progression. 
  \item One can define complex or compositional modes of progression
    in at least two ways. Firstly, one can let a variable $m$ evolve
    in different modes at different periods in its evolution, for
    instance alternating between discrete and continuous update
    windows. The structure of such composition-in-time would be
    characterized by meta-modes. Secondly, one can consider sets of
    variables whose members evolve according to different modes of
    progression, with some sort of coordination between them. An
    example are handshake protocols to
    synchronize different components in asynchronous digital hardware
    platforms \cite{Sparso02}.  Such
    composition-in-space would again be characterized by
    meta-modes. Both ways could also be combined.

\end{itemize}

This indicative list must suffice here. A more complete catalogue of
modes of progression, and an in-depth study of which modes can be
obtained by formal operations from which other ones and for what
formal types of variables and laws of evolution, remain to be worked
out. Such a theory would be needed for a full understanding of our
theme, ``computational extensions of timescales''. Two 
key questions which await a systematic investigation are the following:
\begin{enumerate}
\item How can the loss of information be characterized when one
  abstracts variables to new modes of progression? Can modes of
  progression be ordered in an abstraction hierarchy, such that
  variables with modes higher in the hierarchy can be defined from
  variables with modes lower in the hierarchy, but not vice versa?
  Presumably the physical time mode $t [\mbox{sec}]$ would lie at the
  bottom, and the atemporal logical inference steps of progression
  mode $n$ in symbolic proof engines or the Turing machine would be
  found very high in the hierarchy. 
\item If an abstract computational model is formalized with modes of
  progressions that lie high in the abstraction hierarchy of modes of
  progression, are there systematic ways to successively design less
  abstract models, ultimately arriving at an analytical physical model
  with mode $t [\mbox{sec}]$, such that the more abstract models in
  this sequence can be defined from the ``lower'' ones? This is the
  question of systematic ``compilation'' mechanisms which can bridge
  different levels of modes of progression. In practical development
  of computational models, such down-compilation cascades will rarely
  be driven down to the point where an analytical system model is
  met. Hierarchical ``software'' development will rather build on a
  lowest-level abstract computational model provided by the hardware
  manufacturer, like machine instruction sets in digital computing. To
  construct this root computational model, the hardware manufacturer
  has to build a bridge between an analytical system model and
  computational abstraction. With regards to timescales, on its
  hardware side this bridge must meet all physical timing constraints,
  dynamical stability preserving conditions and the calibration of
  abstract computational measures or mechanisms of timing that are
  offered at this root level (in the digital domain: clocks); and on
  the computational side it must offer such elementary modes of
  progression that allow the software developer to define higher-level
  modes of progression with the greatest possible freedom.

  An example of a specific compilation hierarchy of this kind is the
  neuromorphic system engineering framework proposed by
  \citeA{Zhangetal20}. Such top-down compilation methods would be
  pivotal for engineering practice in order to systematically design
  top-down mapping sequences from the formal task model (Subsection
  \ref{subsecTask2Model}) through a sequence of ``increasingly
  real-time'' abstract computational models down to the analytical
  model. When compiling down to a modeling (or ``programming'') level
  that lies closer to the hardware and whose mode of progression is
  more closely related to physical time $t[\mbox{sec}]$, a twofold
  challenge must be met: the desired computational functionality must
  be preserved in the compiled model, and the lower-level timing model
  must be consistent with an ultimate further compilation down to
  physical time $t[\mbox{sec}]$.
\end{enumerate}

\subsection{How a computing system is (or is not) coupled
  into its environment}\label{secCyberneticAlgorithmic}

A computing system is
necessarily receiving input and generating output --- otherwise we
(the authors) don't consider it as ``computing''. Input can take many
forms (e.g.\ keyboard strokes, sensor readings, ethernet signals,
on-board bus feeds), as can the output (e.g. screen displays, robot
motor commands, signals sent out to wires or antennas). Input and
output couple a computing system to its physical environment. This
coupling establishes important temporal conditions for a computing
system. Many much-used but not unequivocally defined concepts surround
this I/O-coupling of a computing system with its physical context,
like online / offline processing, real-time processing, latency,
anytime algorithms, event-based processing, and more. A computing
system can be temporally I/O-coupled into its physical environment in
many ways.  If we want to get a clear view of the timescales theme, we
must first get a clearer picture of these I/O modes.

\begin{figure}[htb] 
\centering
\includegraphics[width=10cm]{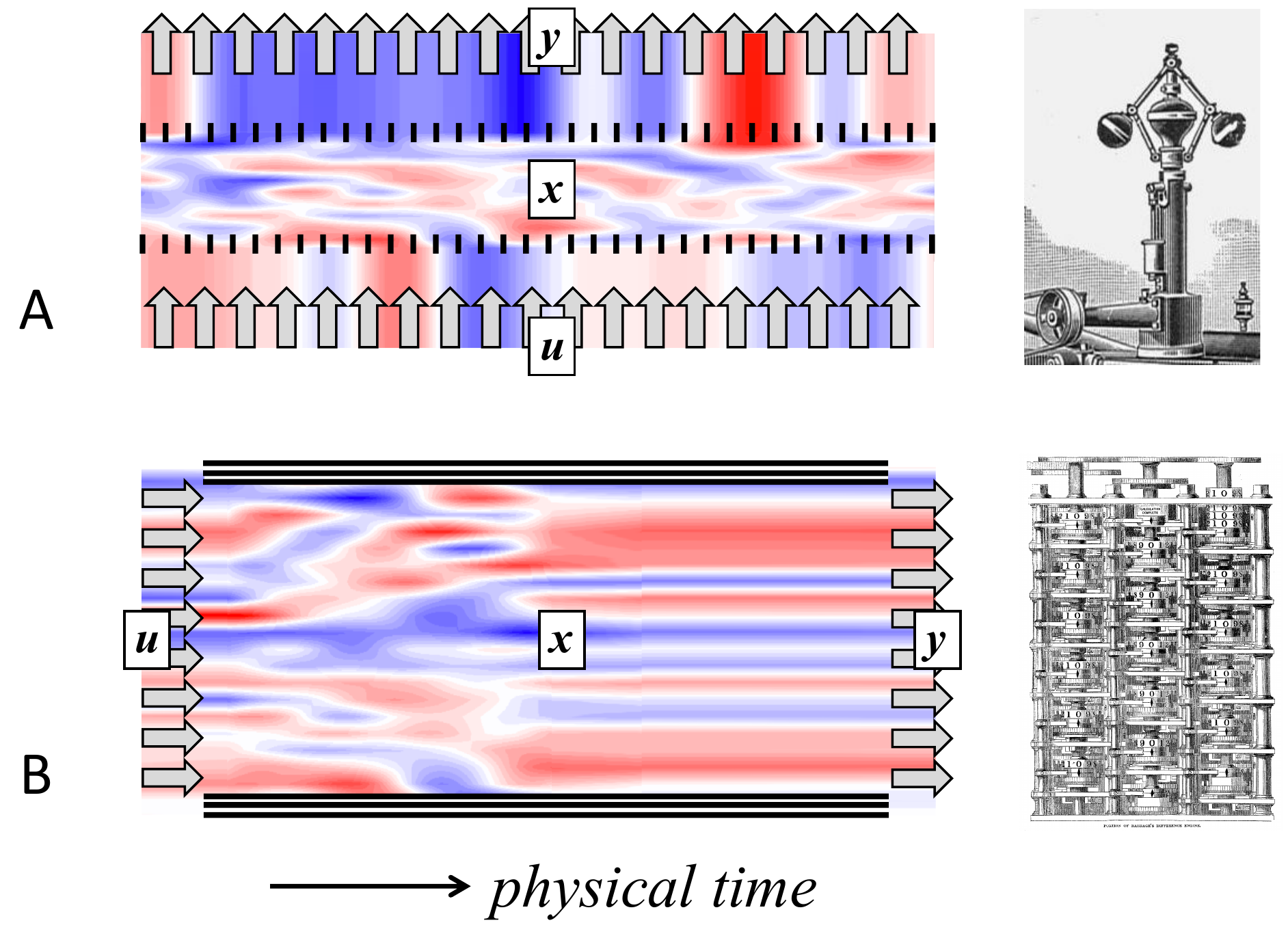}
\caption{Schematic of the cybernetic and the algorithmic mode of
  coupling a physical computing system to its environment through
  input and output. In the cybernetic mode ({\bf A}), the computing
  system with state $\mathbf{x}$ is continuously infused by input
  $\mathbf{u}$ and continuously emits output $\mathbf{y}$ through its
  open boundaries (black striped bands). In the algorithmic mode ({\bf
    B}), input $\mathbf{u}$ effects a setting of an initial system
  state $\mathbf{x}$. During the computation, the system is isolated
  from its environment (black stripe triplets). The autonomous system
  dynamics converges to a stable state which can be read out as output
  $\mathbf{y}$ at any time after stabilization. This requires
  non-volatile memory devices or attractor-based internal dynamics
  where the persisting interal states are attractors (see Section
  \ref{subsecAttractorWM}). Historical drawings
  taken from \url{commons.wikimedia.org}. }
 \label{figtwoMainIO}
\end{figure}

In order to characterize I/O-coupling modes, we propose two basic,
``pure'', complementary modes and characterize others as combinations
and mixtures made from these two basic ones. We will call these two
basic modes the \emph{cybernetic} and the \emph{algorithmic} mode:

\begin{description}
\item[The cybernetic mode] is classically exemplified by Watt's
  centrifugal governor. The computing machine continuously receives
  input $\mathbf{u}$ from the task environment and continuously
  transforms this to continuous output $\mathbf{y}$ through internal
  transformations based on a physical system state $\mathbf{x}$. In
  the case of the governor, the input is physical rotation, the output
  is linear displacement of a steam valve slider, and the
  transformation is effected through gravitation competing with
  centrifugal forces through the geometrical mechanics of
  interconnected shanks (Figure \ref{figtwoMainIO}{\bf A}). Other
  representatives of the cybernetic mode are analog electronic signal
  processing and control systems and the nervous systems of worms and
  insects. The archetypical formal model is the ODE schema
  $$\dot{\mathbf{x}} = f(\mathbf{x}, \mathbf{u}), \;\; \mathbf{y} =
  g(\mathbf{x}).$$ Note that the $\dot{\mathbf{x}}$ refers to the
  state change rate with respect to real physical time $t$; that this
  input-to-output transformation is causal (output $\mathbf{y}(t)$
  does not depend on future inputs $\mathbf{u}(t+h)$); that the
  computation typically has memory ($\mathbf{y}(t)$ is co-determined
  by earlier inputs $\mathbf{u}(t-h)$) and that it may have delay
  ($\mathbf{y}(t)$ is only determined by inputs earlier than some
  $\mathbf{u}(t-d)$). Other formalisms that are naturally used for
  cybernetic models of computing include iterated maps or recurrent
  neural networks.  The ongoing operation is
  \emph{entrained} to the physically evolving input stream. This does
  not mean that if the input would be administered slower or faster,
  the output signal would remain the same except for becoming
  correspondingly slower or faster too --- 
  slower or faster input will typically lead to a new output signal
  that differs form the original one in more ways than temporal
  extension or contraction.  It does not
  matter whether the driving input signal is continuous-time or
  discretely sampled; the important characteristic is that it is not
  interpreted as a sequence of triggering events interspersed with
  waiting / idling times but as a seamless stream.

\item[The algorithmic mode] is classically exemplified by Babbage's
  difference machine, which could compute (among other)
  polynomials. This machine gets its input by an operator
  setting some adjustable number scales. Then the operator starts and
  drives the computing process by manually turning a crank handle
  (with bodily effort, see
  \url{https://www.youtube.com/watch?v=BlbQsKpq3Ak}). The machine
  prints the result's digits on a tape as they become determined. When
  this process is completed, the operator can stop crankhandling and
  read the result. Other ``pure'' representatives of this computing
  mode are electronic pocket calculators and digital computers. A less
  pure (but after some abstraction still pertinent) example is a human
  chess playing novice who knows the rules of the game but otherwise
  lacks experience, and who in order to determine the next move
  simulates in his/her brain a choice of possible move-countermove
  continuations. The archetypical formal model is the Turing machine
  whose transition law is
  $$T: Q \times S \to Q \times S \times \{l, r\},$$
  where $Q$ is the finite set of states of a control automaton, $S$ is
  the finite set of symbols that can be written on a tape, an
  $\{l, r\}$ are the left or right moves of the read/write head on the
  tape. Note that in the abstract Turing machine model there is no
  reference to real physical time --- the next state and symbol are
  \emph{logically} determined by the previous one, not
  physically-causally. When an algorithmic machine is physically
  realized (as in the schematic in Figure \ref{figtwoMainIO}), of
  course it operates in physical time $t$. Characteristic differences
  to the entrained mode of operation of a cybernetic machine are that
  computationally equivalent algorithmic machines can be physically
  built which operate slower or faster with regards to physical time;
  that the input is set only at the beginning of operation by fixing a
  starting state; that during its operation an algorithmic machine is
  isolated from its environment; that its dynamics converges to a
  stable state which can be read out as output after any physical
  waiting time after convergence.

  Digital computing architectures, operating systems or programs can
  be designed according to an \emph{event-based} paradigm. This term
  may mean different things in different contexts, but the general
  idea is that the entire system or subsystems do nothing until they
  are activated by an incoming trigger event with fresh input
  information, then autonomously carry out a computation at which end
  they emit triggers/output, which is sent to other subsystems or the
  user. An accompanying idea is that there is no general scheduler
  meta-mechanism: event-based processing is often considered as
  asynchronous, assuming that when a subsystem has terminated its
  triggered process, its output is immediately sent off to trigger
  other subsystems. A clear example are implementations of spiking
  neural networks where spikes are sent off, transmitted, and received
  asynchronously with an address-event representation (AER) protocol
  \cite{Mostafaetal15b, IndiveriCorradiQiao15}. The classical
  mathematical model of event-based processing are Petri nets
  \cite{Petri62}. If one views the term ``event-based'' broadly, all
  systems that we call algorithmic can be regarded as event-based.  On
  the hardware level, at the finest-grained resolution each Boolean
  gate may be triggered anew in every clock cycle. Abstract
  computational models (typically computer programs) can be seen as
  event-based in several ways, where the background intuitions differ
  with the programming style. In imperative programming, each
  assignment of a value to a variable $X$ is an event which sets the
  state of this ``subsystem'' $X$ to the new value until it is changed
  by another write event. If functions are defined in an imperative
  program, they can be seen as subsystems that remain inactive until
  ``called'', and after having finished their called operation they
  revert to an initial state. In object-oriented programming, objects
  are the subsystems whose internal state remains stable after its
  input-triggered internal operations have terminated and the object
  is accessed again.

  Above we characterized the cybernetic mode as having the computing
  system 'entrained' by the input. In contrast, the algorithmic mode
  could be characterized as \emph{autonomous} in the sense that while
  it is executing one of its subroutines, the respective processing
  system is shielded against input.
  
\end{description}

The cybernetic and the algorithmic modes will often be realized in
mixed forms in real computing systems. For instance, online processing
tasks realized in digital real-time operating systems generate input
responses that are closely coupled to the input data stream at some
temporal resolution; however, this reactive behavior is achieved by
running classically algorithmic subprocedures at a rate that is fast
enough to always terminate before the next relevant input data point
arrives. A case in point are reactive robot motor control programs
where the sensory input is sampled at a high frequency, with motor
commands computed and updated at the same frequency. More involved
mixtures of cybernetic and algorithmic modes have been discussed in
cognitive science. For example, \citeA{Botvinicketal01} examines how
humans monitor conflicts in their automated, ``habitual'' processing
(cybernetic mode) and then launch conflict control
procedures (algorithmic).

The analytical system models in our schema from Figure
\ref{IOcompleteVsNot} are always entrained to their input signals
because the physical systems, which they model, are continuously
driven by their real-time input signal. Algorithmic processing
episodes can only be defined in abstract computational models, which
offer the option of defining computational state sequences that are
decoupled from the real-time state evolution in the underlying
physical system model. Yet, this decoupling is not absolute. While an
algorithmic state evolution is going on in some computational
subsystem, new physical input must be either ignored, or buffered for
future processing, or taken care of by other computing subsystems in
parallel. All of this is connected with core questions that arise in
real-time, anytime, and parallel computing. We cannot attempt to
explore the resulting intricacies here.

  \subsection{Timing of input  and output}\label{subsecInputtime}

Input can be delivered to an abstract computing model in
various ways, for instance
\begin{enumerate}
\item as (part of) the \emph{initial state} as in Turing machines, the
  Hopfield network, or the abstract model of a computing system
  proposed by \citeA{Horsmanetal14} --- this is constitutive for what
  we called the algorithmic mode of processing above,  
\item as \emph{clamped} input like in the Boltzmann machine
  \cite{Ackleyetal85}, or
\item as \emph{interactive} input which is given to the computing
  system intermittently during its operation by a user, as in models
  of interactive Turing machines \cite{vanLeeuwenWiedermann01}, or
\item as an online \emph{signal} which is continually fed to
  the computing system during its operation.
\end{enumerate}
It is possible to formally represent the first three of these options
as special cases of the last one, for instance as follows:
\begin{enumerate}
\item The input signal gets an extra channel with only two possible
  values 1 and 0, which is set at start time $t=0$ to 1, and in
  addition a constant signal that at every time is equal to (an
  encoding of) the
  desired system start state $\mathbf{x}(0)$, and add a special
  mechanism to the system equations which set the system state to
  $\mathbf{x}(0)$ when the extra input channel reads ``1''.
\item The input signal has the constant ``clamped'' input value at all
  times.
\item Like in 1., add an extra 0-1 indicator input channel to an input
  signal which in its other channels has a ``payload'' input sequence, with the
  system equations configured such that the payload input is read
  whenever the indicator channel shows a ``1''. 
\end{enumerate}
These methods to cast inputs of sorts 1.--3. as signals have been
used, for instance, in the extensive studies of long-term memory
capabilities in recurrent neural networks in \citeA{MartensSutskever11}
and \citeA{Jaeger12}.

In the remainder of this report we will thus consider ``inputs''
always as a temporal signal that is fed to the computing system
throughout its operation. 

By analogous arguments we will consider ``outputs'' likewise as
temporal signals that are issued by the computing system throughout
its operation.

\subsection{Concepts of timescales}

So far we worked out what we understand by a computing system, at a
level of differentiation that will be suitable for discussing
computational extensions of timescales. But we have not yet
clarified what we mean by ``timescales''. In the following subsections
we take a closer look and will find that there are quite different
sorts of them.

\subsubsection{Next to the physical basis: timescales as time
  constants}\label{subsecTimeConstants}

Computing systems are physical systems. Arguably the most popular
formalism in physical system modeling is ordinary differential
equations (ODEs). Physicists, neuroscientists or electronic circuit
engineers will almost by reflex describe their target systems through
state vectors $\mathbf{x}(t) \in \mathbb{R}^n$, where the individual
system variables $x_1, \ldots, x_n$ correspond to physical quantities
measured in standard physical units like Volt or Ohm, and the dynamics
of the system is expressed by  coupled differential equations
\begin{equation} \label{eODEbasic}
  \tau_i\; \dot{x}_i =
  f_i(\mathbf{x}, \mathbf{u}, \mathbf{a}),
\end{equation}
where $i = 1, \ldots, n$ and $\mathbf{u}(t)$ is an optional input
signal and $\mathbf{a}$ is an optional vector of control parameters,
and $\tau_i$ is the \emph{time constant} of the dynamics of the
physical quantity $x_i$. When $\tau_i$ is large, physicists and
dynamical systems mathematicians call $x_i$ a slow variable, and when
it is small, a fast one. When all these equations are set up
appropriately, the model will describe the temporal evolution of all
the physical quantities modeled by the variables $x_i$ in numerically
correct rates of change with respect to the real-world physical time
$t$, standardly expressed in seconds. One sometimes calls the time
constants $\tau_i$ suggestively the \emph{native} or \emph{intrinsic}
timescales of the respective physical quantities, with a background
intuition that these time constants reflect the ``real'', or
``causal'', or ``physical'' timescale of the concerned quantity. We
will use the term ``physical time constant'' in a quite narrow sense,
namely for time constants associated with the physically measurable
variables in ODE models whose variables are quantified in standard
units of measurements.

We note that the absolute values of such time constants depend on the
choice of units of measurement, both for system state variables
and for time itself. If one measures time in hours instead of seconds,
all time constants will have to be scaled by 1/3600; if one measures
some voltage $x_i$ in mV instead of V, one has to scale its time
constant by a factor of 1000. The absolute values of
time constants, and the ratios of time constants of two variables that
correspond to different physical sorts (like voltage vs.\ current),
are thus in essence arbitrary, and it makes little sense to speak of a
fast variable only because its formal time constant is small.

There is however a crucial aspect of physical time constants that is
not arbitrary. It comes to the surface when one compares two 
models A and B of a physical system, both expressed in the same
units of measurement, where between the two versions there
is a 1-1 mapping of a subset of system variables. For instance, in an
electronic circuit model A one may change one resistance to obtain version
B, which would give a global 1-1 variable mapping between A and B. Or
in A one might add a little extra subcircuit to get B; the 1-1 mapping
would be between the variables of A and all the variables in B that do
not belong to the newly inserted subcircuit. When one compares A with
B, the ratios between the time constants of the 1-1 mapped system
variables is independent of the choice of units of measurement, and it
is these ratios which guide a system engineer to let the final design
fulfil its temporal specifications.

There are of course other modeling formalisms for physical systems
besides ODEs, for instance partial or stochastic differential
equations or more general sorts of stochastic process formalisms. In
some of these one can find canonical correspondents to ODE time
constants, for instance for the drift component in stochastic
differential equations (chapter 15 in \citeA{Kuehn15}), or the
(inverse of the) variance of a Gaussian transition kernel in
continuous Markov processes. We are however not aware of a unified
definition of time constants across several classes of modeling
formalisms, and will restrict our discussion to ODE models since these
are so widely used, and since the mathematical theory of what
mathematicians call \emph{multi-timescale systems} is rooted in ODE
models \cite{Kuehn15}. 

A hallmark of physical time constants is that if one wants a system in
whose model they are changed, one has to change the physical make-up
of the system. An electronic circuit engineer who wants a certain time
constant become faster or slower will have to create a new physical
variant of the system, for instance by changing 
resistances or adding new circuitry. 

In some scenarios it may be possible to change physical time constants
at runtime through methods of online hardware configuration. This
option is increasingly used in scaled digital systems where the
processor is subdivided into ``islands'' (typically cores in multicore
chips) that each have their
individual voltage and/or clock frequency settings which can be
dynamically controlled for energy efficiency and thermal management
\cite{Herbert07, Parketal13}. Such
online-changeable time constants would be mathematically modeled with
ODEs of the kind
$\tau_i \dot{x}_i = c_i(\mathbf{x}, \mathbf{u}, ...) \cdot
f_i(\mathbf{x}, \mathbf{u}, ...)$ where the factor function
$c_i(...) > 0$ represents the online time constant modulation of this
equation, yielding an effective time constant of $\tau_i / c_i(...)$
for the ODE $\tau_i \dot{x}_i = f_i(\mathbf{x}, \mathbf{u},
...)$. This makes mathematical sense if $c_i(...)$ changes its value
much slower than $f_i(...)$.

Physical system models made by physicists or engineers are almost
always intended to be \emph{analytical} models, as opposed to the
blackbox models that are typical in machine learning. In an
analytical modeling spirit, the model's variables and formal dynamical
laws should correspond to the  physical quantities and
physical mechanisms in the target system. Thus, the model variables
$x_i$ should correspond to classical physical quantities which are
measured with standard physical units. As a consequence, an
analytically minded physical system modeler cannot do what
mathematicians often like to do, namely investigate the same system in
a transformed coordinate system. For instance, in a two-dimensional
system model with $x_1$ capturing a voltage in Volts and $x_2$ a
resistance in Ohms, applying a linear coordinate transformation $A$ to
state vectors $(x_1, x_2)'$ would give new system variables
$(z_1, z_2)' = A\; (x_1, x_2)'$ which would formally correspond to
weighted mixtures of voltages with resistances, for which there is no
meaningful physical unit of measurement, and it would not be possible
to build a measurement apparatus that could directly measure $z_1$ or
$z_2$. Analytical ODE system models and their time constants are in a
deep sense ``locked'' to a specific coordinate system.

An intuitive interpretation of physical time constants is not
straightforward. Making a time constant $\tau_i$ smaller or larger in
a system model does not imply that the concerned system variable $x_i$
always changes faster or slower. For example,
\begin{itemize}
\item when a system whose time constants are all very small (hence
  called ``fast'') is close to a fixed point attractor, and its
  input signal is constant (or this system has no input), the system
  variables will be asymptotically coming to a standstill despite
  their fast time constants;
\item a system whose model has all very large (``slow'') time
  constants can exhibit fast oscillations even with absent or constant
  input if its internal feedback gains are strong enough;
  \item a coupled oscillator system model with a ``fast'' variable
    $x_i$ that is associated with a small time constant $\tau_i$ may
    display slow changes of $x_i$ first, which become fast
    oscillations when $\tau_i$ is \emph{increased} --- due to shifting
    the overall dynamics to a resonance frequency which was suppressed
    when $\tau_i$ was small. 
\end{itemize}
There is thus not a universal connection between making time constants
smaller (in model simulations and/or by modifying the physical system)
and observing that the concerned system variables ``move faster''.

It requires a case-by-case discussion to understand how a specific
physical time constant impacts on dynamical ``speed''
properties of the modeled system. The observed ``speed of change'' of
a variable $x_i$ with time constant $\tau_i$ will depend, among other
factors, on fast/slow properties of the driving input, strength of
system-internal feedback, attractors present in the system, or whether
the system is observed in transients or close to attractors. Intuitive
or mathematical timescale-relevant interpretations of  
$\tau_i$ will differ. Examples for such case-by-case interpretations
of time constants:
\begin{itemize}
\item When a system is driven by slow input, such that its (single)
  fixed point attractor is adiabatically following the input, the time
  constant $\tau_i$ characterizes an exponential \emph{rate of convergence}
  to the fixed point, which for changing input translates to \emph{more
  precise} or \emph{shorter delay} input tracking when  $\tau_i$ gets smaller.
\item (Only) in system models without input, scaling all time constants
  by the same scaling factor $a$ will make the system trajectory
  $\mathbf{x}(t)$ move slower or faster through the state space
  $\mathbb{R}^n$ in proportion to the scaling factor $a$. This may be
  the original motivation to call the $\tau_i$ by the name ``time
  constants''.
\item In singular perturbation methods in mathematical studies of
  so-called slow-fast systems, the ODE models have two groups of
  equations, one group with small and the other with large time
  constants (the ``fast and slow subsystems''). When the ratio of the
  small over the large time constant approaches zero, specific
  operating and interaction conditions for the two subsystems can be
  isolated:
  \begin{itemize}
  \item The fast subsystem approximately behaves as if the slow
    subsystem is standing still, yielding a control parameter vector
    to the fast subsystem. This admits bifurcation analyses of the
    fast subsystem.
  \item The dynamics of the slow subsystem can be studied using only
    the equations of the slow subsystem on the \emph{critical
    manifold} $C_0 \subset \mathbb{R}^n$ embedded in the total state
    space, where $C_0$ is the set of zeros of the fast
    subsystem. Close to $C_0$, system trajectories will be ``pulled
    along'' these slow trajectories within $C_0$ if $C_0$ is an
    attracting surface. 
  \end{itemize}
\end{itemize}

It becomes clear from such considerations that time constants in
physical system models capture ``real'' physical dynamical
characteristics of the concerned state variables on the one hand, but
that on the other hand these physical characteristics do not 1-1
translate to phenomenal ``fastness'' or ``slowness''. The time
constants which one finds in analytical ODE models of
physical systems (or rather, their reciprocals $\tau_i^{-1}$)
would maybe better be called ``coupling strength constants'' than ``time
constants'': the larger  $\tau_i^{-1}$, the stronger is $x_i$ impacted
by the other system variables on the r.h.s.\ in $\dot{x}_i =
\tau_i^{-1}\,f_i(\mathbf{x}, \mathbf{u}, \mathbf{a}).$ Since an
analytical system is a stand-in for the real physical system, and its
dynamics reflect physical causality, one could also say that these ODE
time constants model \emph{causal} effects.

\subsubsection{Timescales of change} \label{subsecTimesOfChange}

In contrast to the causal effects that are reflected in an analytical
system model, ``timescales'' in abstract computational models are
\emph{phenomenal}. When we speak of timescales in the context of
abstract computational processes, we describe \emph{how} things
change, not \emph{why}.  

In this subsection we consider the phenomenal aspect of ``speed of
change''. This is maybe the aspect that is most immediately associated
with the word ``timescales''.  The core intuition here is that we call
a temporally evolving variable ``fast'' if it changes strongly within
short time intervals, like high-frequency oscillations; and we call it
``slow'' if it changes only a little or not at all as time goes on.

This basic aspect of ``changing fast'' vs.\ ``changing slowly'' is an
idea of \emph{motion}, ``fast'' meaning that much \emph{distance} is
covered per time unit. We can
thus discuss timescales of change only for mathematical objects that
lie in spaces where some kind of  distance measure is available, like
a mathematical metric or the
Kullback-Leibler divergence. This includes  objects like
\begin{itemize}
\item points in metric spaces,
\item suitably restricted classes of functions into metric spaces like
  $f: D \to \mathbb{R}^n$ which yield metrizable function spaces;
  this includes many kinds of vector fields,
\item probability distributions over measurable spaces,
\item nodes in an undirected graph where distance is graph distance,
\end{itemize}
and more. We leave it at that, and will refer to mathematical objects
that come with some notion of distance as \emph{metric variables}
 with the understanding that we admit other distance-like
measures besides the standard mathematical definition of a metric. 

In many important formal models
of computing systems (including the Turing machine), the three sorts
of formal objects $\mathbf{u}(t), \mathbf{x}(t), \mathbf{y}(t)$ are
however not metric variables but discrete objects like truth values,
symbolic expressions, graphs or neural spikes. In order to start
discussing ``speed of change'' in such situations, one can attempt to
create meta variables which are metric in some sense and thus admit
the definition of rates of temporal change. A common example is to
define spatial or temporal averages of spike counts to get
continuous-valued spike rates $m(t)$ which are metric variables. In
a run of a Turing machine with tape alphabet $\{A, B\}$ one could (an
entirely arbitrary example) define as a metric variable $m(t)$ the
the number of $A$'s written on the tape during the last 100 update
cycles up to time $t$. This would make $m(t)$ a number between 0 and
100 which would change by at most $\pm 1$ in a step from time $t$ to
$t+1$, a measurable increment upon which timescale characteristics can
be defined.

Thus, if we have a time series $v(t)$ or $v(n)$ in continuous or
discrete time, and $v$ is a metric variable --- how can we define
and measure ``speed of change''? Here are some important options.

\begin{itemize}
\item A signal processing engineer will transform $v(t)$ or
  $v(n)$ to the frequency domain and call it slow when the Fourier
  spectrum is dominated by low frequency components (which in turn
  leads to the question how, precisely, one will declare frequency
  spectra to be dominated by what frequency components).
  \item In wavelet transforms, slow components of $v(t)$ or
  $v(n)$ could be associated with the projections on wide-support wavelets.
\item A mathematician may want to compute norms like
  $1/T \;(\int_{t = 0}^T |\dot{v}(t)|^k \, dt)^{1/k}$ and call $v$ fast
  if this norm is large, on the grounds that a large norm implies ``a
  lot of change'' during a reference time interval $T$.
\item Such norms however would also be large for a constant ramp
  signal $\dot{v}(t) = \mbox{const}$ when the slope is large. This may
  often not be what one wants, and one might prefer higher-order
  measures of change, like $1/T \;(\int_{t = 0}^T |\ddot{v}(t)|^k \,
  dt)^{1/k}$ which quantify \emph{variations} in speed of change
  instead of only speed of change itself.
\item All such norm-based measures will scale linearly or nonlinearly 
  with scalings $\alpha \, v$ of  the concerned signal. This
  may or may not be desired, and if not, the signal $v(t)$ would first
  have to be normalized in some way.
\item A complex systems researcher might consider $v(t)$ as slow when
  its autocorrelation plot has a ``fat tail'', that is when there are
  long-range temporal correlations which decay slower than
  exponentially. More generally, autocorrelation spectra may be said
  to contain slow components when autocorrelation values for large
  timelags are large.
\item A dynamical systems mathematician will identify regions in an
  ODE state space $\mathbb{R}^n$ where $\mathbf{v}(t)$ moves slowly
  vs.\ regions where it moves fast, measured by
  $| \dot{\mathbf{v}} |$. Here slowness/fastness becomes a local
  property that changes with time $t$. 
\end{itemize}

This list illustrates that there are many natural options to define
quantitative measures of rates or speeds or variations of change. The
use of norms vs time-spectral transformations vs other alternatives
involves a complex trade-off between many factors. It will depend on
the application at hand and the objectives of the designer.

In all these measures there is another source of freedom of
definition, arising from the question how locally vs.\ globally one
defines these measures. Consider
a signal $v(t)$ which jumps between the values 0 and 1 with steep
flanks, staying at the two extreme values for extended periods. To the
degree that the flanks are steep, most of the measures listed above
will give large values in small measurement intervals around the
flanks, and zero ``speed of change'' when measured within one of the
two plateaus. One might opt for defining the speed of change for such
a signal (if it is stationary in the sense of stochastic processes) as
the measure value in the limit of the measuring interval $T$ going to
infinity. But this will hide the possibly interesting and relevant
fact that there are clearly discernible fast and slow
subperiods. Measures of speed of change should thus be qualified by
the observation interval. Instead of a single speed-of-change
measurement value for a signal $v(t)$ one gets a hierarchy of such
values, with levels ordered by durations of $T$; and within each level
$T$ one gets a distribution over measurement values.

Summarizing, a full account of analysing timescales of change for a
stationary signal $v(t)$ or $v(n)$ would include
\begin{enumerate}
\item the choice of a numerical measure $\mu$ that yields a ``speed of
  change'' value for any given observation window $[t, t+T]$,
\item for each $T > 0$, a probability distribution over the outcomes
  $(\mu([t, t+T]))_{0 \leq t \leq \infty}$.
\end{enumerate}

\subsubsection{Timescales of reactivity}\label{subsecReactivity}

For computing systems it is relevant to know \emph{how soon will the
system react to inputs?} This question comes a number of
interesting variants:

\begin{itemize}
\item In the classical model of symbolic computing, based on
  formalisations of algorithms like the Turing machine or computer
  programs, the question of reaction time to an initial input has
  become thoroughly studied in the theory of \emph{computational
    complexity}, in particular \emph{time complexity}. The time
  complexity of an algorithm is defined via the number of discrete
  state update steps that the algorithm needs to return the result,
  defined as the asymptotic growth function by which this
  step-counting time increases when the size of the input argument
  grows.
\item Tasks that require online responses to event-like input from the
  user often come with an objective to react to the inputs with short
  latency.
\item In neuroscience modeling and analog signal processing one
  considers the real-time \emph{delays} after which the hardware
  system shows a response to information in the continuous input
  signal. Different characteristics of the response
  (e.g. accuracy or statistical reliability) can reveal themselves
  after different delays, whith more accurate or more reliable
  response components arriving later. Similarly, classical symbolic
  \emph{anytime algorithms} can be queried for output after different
  processing durations, earning more precise or reliable results when
  one waits longer.
\item In feedback tracking control systems, higher feedback gains lead
  to faster and more precise tracking, at the risk of coming closer to
  instability of the control loop.
\item Time-to-failure prediction systems are often used in predictive
  maintenance of engines, generators or other industrial
  systems. These algorithms are typically machine learning
  models. Here a fast reactivity would mean that \emph{early} warnings
  are generated, which in turn means that the monitoring system reacts
  very \emph{sensitively} to changes in the monitored diagnostic
  signals. While high sensitivity of an online computing process is
  not the same as \emph{fast} reactivity, the desired \emph{early}
  reactions are a closely related aspect of reactivity timescales.
\item In digital signal processing, the clock cycle time sets a lower
  bound on the possible reaction time.
\end{itemize}

  Reactivity can be discussed in relation to the two modes of
  computing that we called cybernetic and algorithmic
  (event-based) in Section \ref{secCyberneticAlgorithmic}. 
  
  In cybernetic computing systems, the reactivity theme surfaces when
  one analyzes the reaction time needed until the desired response
  information appears in the output signal. Signal travel delays play
  an important role. In linear signal processing (continuous time or
  discretely sampled time), an indicator of reaction delays is the
  delay of an impulse response. How this delayed-reaction time is
  defined and measured in general depends on how the processing task
  is defined. For instance, the output may quickly yield an encoding
  of a preliminary but inaccurate response, followed after more time
  by response information of higher accuracy. In multimodal tasks,
  output components of different modalities may arrive with different
  delays.

  In algorithmic computing, reactivity questions arise when real-time
  constraints are imposed by the task. Subsystems must finish
  their internal processing within an allotted physical timespan
  $\Delta t [\mbox{sec}]$. Real-time operating systems may be needed
  to provide a universal computational infrastructure which admits
  task programming under physical time conditions. Designing such
  systems on the hardware as well as on the abstract computational
  model must meet challenges of resolving conflicts between competing
  resource demands, synchronization, and varying computational loads
  (data throughput demands, variable input sampling rates, variable
  computational complexity of current subtasks).

\subsubsection{Timescales of memory} \label{subsecMemTimescales}

For a dynamical system to ``compute'' it is quintessential that it can
preserve and transform information over time --- that it can memorize.

The most natural (if not naive) concept of ``memorizing'' is to cast
it as ``storing'': putting things on a shelf where they can be fetched
whenever needed. This is in fact the concept of memory that permeates
digital computing practice and theory: ``writing'' bits into
non-volatile memory devices and ``reading'' or ``deleting'' them when
needed. 

But in many non-digital hardware systems including brains,
non-volatile memory devices are not available or are plagued by
numerical imprecision, process variation (at fabrication time),
temperature sensitivity, stochastic fluctuations and device
mismatch. The spectacular long-term studies by psychologist
\citeA{Bartlett32} highlight that human memory systems do not
``store'' information but keep on transforming any memorized
information, continually re-shaping and re-coding memory traces in
order to preserve an overall consistent cognitive world
representation.  ``Memorizing'' thus is a complex game which includes

\begin{itemize}
\item \emph{encoding} information in dynamical sub-processes at the
  time when this information appears in the input or when it was
  internally generated,
\item \emph{propagating} this encoding during some time lag by keeping
  traces or transformations of it ``alive'' in the ongoing
  system dynamics,
\item and \emph{decoding} it back to a useful format
  when later needed.
\end{itemize}

Each of these stages can become more concretely spelled out and
realized in a multitude of ways. They differ with regards to, for
example, how the elusive concept of ``information'' is understood in
the first place; whether the input is given only once at the beginning
of a run of an algorithm or streams in continuously (and similarly
whether the decoding is done only once at the end of a run or a
continuous stream of readings from memory is needed); whether a
gradual degradation of the (encoded) information over time is
acceptable or not; what kind of dynamical mechanism is exploited for
each of the three stages; or whether the memorizing dynamics are
interwoven with learning processes that change the processing system
itself (arguably always the case in biological brains), etc. Further
facets are added to the picture by the overlay of different memory
mechanisms in the same processing system to cope with multiscale
temporal tasks; meta-mechanisms of \emph{attention} to decide which
parts of currently available information needs to be memorized; other
meta-mechanisms of \emph{addressing} for determining where information
can be found to be decoded; and not to forget: \emph{forgetting} or
overwrite mechanisms to free ``memory capacity''.

There are many options for mathematical definitions and numerical
measures of ``how much information of what sort'' is propagated
through a time lag $\Delta$. In the reservoir computing literature, a
popular measure is the \emph{memory capacity} introduced in
\citeA{Jaeger02b}. This is a simple measure based on the correlation
between input and output signals which is easily  measured in computer
experiments. However, a more fundamental and insightful
definition/measure would be to define and quantify ``information
transfer'' by the \emph{mutual information} between system states
$\mathbf{x}(t)$ and $\mathbf{x}(t + \Delta)$ separated by a lag
$\Delta$. This measure is independent of encoding and
decoding operations; it is a characteristic of the system
states $\mathbf{x}$ themselves.

This
information-theoretic line of thinking about ``memory'' implies that
one cannot determine whether a system is capable of memorizing by
looking only at a single example pair
$(\mathbf{x}(t), \mathbf{x}(t + \Delta))$: even when one observes that
a system is exactly in the same state
$ \mathbf{x}(t) = \mathbf{x}(t + \Delta)$ before and after the lag,
this does not mean that the system has memorized $ \mathbf{x}(t)$. In
an information-theoretic view, memory mechanisms must be defined and
identified by considering \emph{distributions} of earlier-later system
state pairs, in order to assess whether or how much information has
been transferred through the lag interval.

Or, alternatively to considering distributions over states, one can
define mutual-information-based memory for states that encode
distributions. If $\mathbf{x}(t)$ and $\mathbf{x}(t + \Delta)$
each represent probability distributions, the mutual information between
even a single such pair is defined. Such states-as-distributions
occur, for instance, when $\mathbf{x}(n + \Delta)$ is the probability
vector obtained from in a discrete Markov chain with transition matrix
$M$ by $\mathbf{x}(n+1) = (M')^\Delta\, \mathbf{x}(n)$, or when
$\mathbf{x}(t)$ describes the evolution of a Fokker-Planck equation.

Hybrids between these two options can also be defined. Consider a
stochastic system with states $\mathbf{x}(n)$ or $\mathbf{x}(t)$ whose
states are ``just states'', not representing probability
distributions, for example the states $s \in S$ of a discrete
Markov chain $X(n)$. Every state $s\in S$ can be
associated with the conditional distribution $Y_{s,k}$ over $k$-step
continuations of the process, that is the distribution
over $S^k$ given by
$$Y_{s,k}(s_1, \ldots, s_k) = P(X(n+1)=s_1, \ldots, X(n+k) = s_k \; | \;
X(n) = s).$$ Then, when one has a pair of states
$\mathbf{x}(n) = s, \mathbf{x}(n + \Delta) = s'$, one can define and
measure the information carried from $\mathbf{x}(n)$ to
$\mathbf{x}(n + \Delta)$ by the mutual information
$I(Y_{s,k}; Y_{s',k})$. This way (which can be very much generalized)
of associating or even identifying physical system states with the
conditional distribution of the process' future after that state has a
venerable history in physics \cite{Zadeh69} and has become central in
certain representations of stochastic processes, such as epsilon
machines in complex systems studies \cite{ShaliziCrutchfield01},
multiplicity automata in theoretical computer science
\cite{Schuetzenberger61}, observable operator models in machine
learning and mathematical theory of stochastic processes
\cite{Jaeger98b}, and predictive state representations in
reinforcement learning and agent modeling
\cite{LittmanSuttonSingh01}. A unifying review is given by
\citeA{ThonJaeger11}. This view is also constitutive for the
\emph{predictive brain} hypothesis in cognitive science
\cite{Clark13} which posits that neural brain states encode
expectations about what is going to happen next.

Unfortunately, numerically estimating the mutual information from
samples of two distributions is only approximate, computationally
expensive and becomes practical only when one makes additional
assumptions about the distributions. In practice one will often take
resort to correlational measures which sometimes provide qualitatively
similar insights as mutual information \cite{MetznerKrauss21}.

Defining and measuring from samples the information transfer between
two systems or between two times in the same system has been
extensively studied. Many definitions and estimation algorithms have
been proposed besides basic correlation and mutual information, for
instance \emph{transfer entropy} \cite{Schreiber00} which is an
information theoretic measure, or \emph{Granger causality} which in
its original formulation is correlation-based \cite{Granger69}, or
\emph{correntropy} \cite{Xuetal08} which is a hybrid. The last
reference also gives a brief survey of other measures. Some of these
measures are symmetric (correlation, mutual information and
correntropy), which is maybe counter-intuitive: one may wish that
information be carried forward through time is not the same as the
information ``reflected backward''.  Transfer entropy and Granger
causality are nonsymmetric and have been designed to capture a
direction of \emph{causation}. We cannot attempt a comprehensive
account here.

\begin{figure}[htb] 
\centering
\includegraphics[width=10cm]{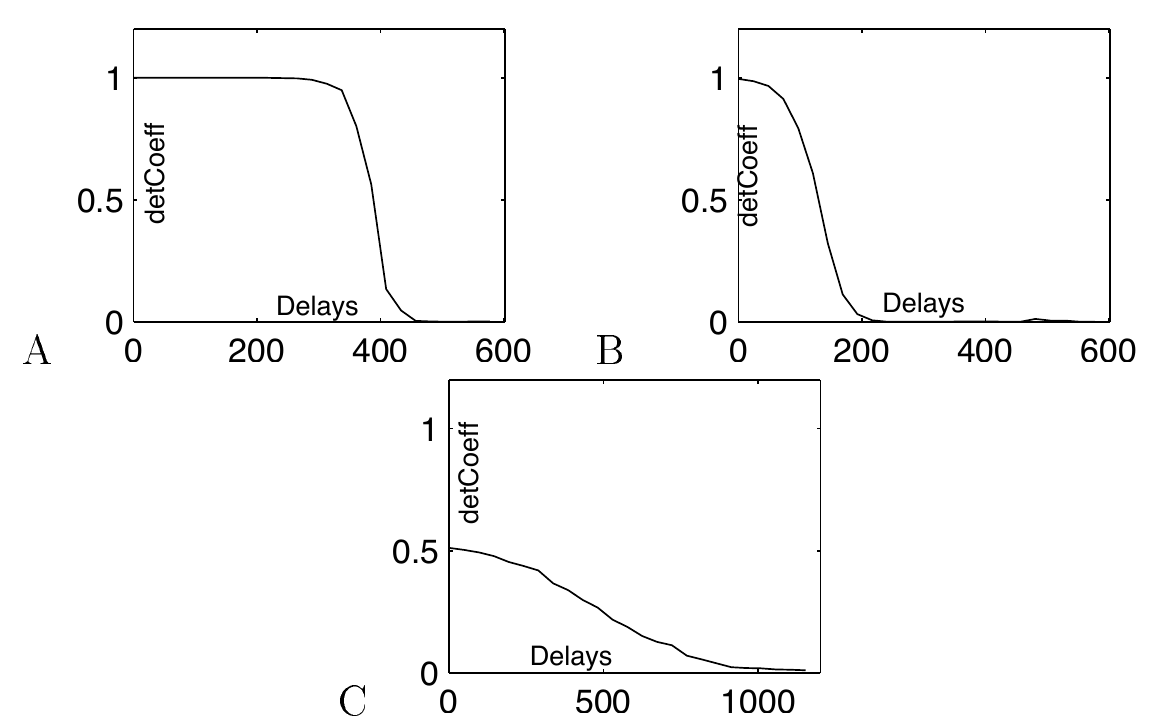}
\caption{Illustrating the diversity of memory curves: Three memory
  curves, showing a correlational measure (determination coefficient)
  between inputs and outputs of the same recurrent neural network
  under the influence of three different hyperparmetrizations. Figure
  taken from \protect\citeA{Jaeger02b}. }
 \label{figMemCurves}
\end{figure}

When one has settled for a measure $M(\Delta)$ of information transfer
across a time lag $\Delta$, in order to define timescales of memory it
is not enough to consider the information transfer across a single
such lag. Instead one should base timescale discussions on
\emph{forgetting curves}
$f: [0, \Delta_{\mbox{\scriptsize \sf max}}] \to \mathbb{R}^{\geq 0},
\;\; \Delta \mapsto M(\Delta)$. These forgetting curves may take
quite different shapes, indicative of different kinds of
forgetting/memorizing. They can range from almost rectangular curves
indicating perfect recollection up to a critical lag after which
there is total forgetting (Figure \ref{figMemCurves}A), to gradual
long-time decay of memory traces (Figure \ref{figMemCurves}C), with
interesting special cases like exponential decay or fat tails. It
becomes clear that speaking about a ``memory timescale'' needs a
careful definition of which measure of information transfer is used
and which are the geometric properties of the forgetting curve.

Many online processing tasks require
the integration of input information across several timescales. The
paradigmatic example is speech recognition. Correctly recognizing the
current sound input in the presence of noise needs disambiguation in
the context provided by earlier input. Relevant previous information
of different sorts stretches back through all levels of the linguistic
hierarchy, from phonemes, syllables, words, phrases, sentences to
texts. Similarly extended memory hierarchies are called upon in many
other applications, for instance navigation and decision-making of
autonomous robots, \cite{NishimotoTani09}, gesture recognition
\cite{Jungetal15}, or predicting chaotic dynamics
\cite{Chattopadhyayetal20}. To address this multiscale memory
challenge, there are several standard approaches in machine learning
models based on RNNs: transformer networks (in deep learning),
hierarchical network architectures where higher model layers have
increasingly longer time constants (in agent modeling), or the use of
large networks (possibly with a population of neurons that have
different time constants; in reservoir computing). Implementing such
models in digital RNN emulations presents no difficulties, because
arbitrary memory spans can be realized by buffering encodings of
earlier input, which in turn is possible due to the presence of
perfectly non-volatile memory devices. However, when only volatile
memory devices are available, as in many analog neuromorphic and other
unconventional hardware bases, realizing such memory hierarchies would
certainly be facilitated if the hardware system offered a wide
spectrum of physical time constants (though this is not the only
option, see Section \ref{secManagingMemory}). Here we want to mention
ongoing research in the MemScales project \cite{MemScales22}. In this
project, a range of novel memristive and other materials and devices
are explored with a focus on their memory timescale ranges and
tunabilities. Among them are filamentary memristive devices like OxRAM
and CBRAM, phase-change based materials as used in PCMs. These devices
and materials give rise to very long memory times, which are
approaching non-volatility. However, these materials come at the price of a more limited  tunability.  It is
difficult to modulate these devices or materials in a way that a wide
range of memory times is achieved. Only 2 -- 3 decades are within
reach. However, we also study a special class of amorphous materials
that exceptionally low leakage currents in the attoampere range and
even below. They can be fabricated as TFTs (thin-film transistors)
with materials such as amorphous Indium Gallium Zinc Oxide
\cite{Miglioratoetal15}. When the proper control circuity is added to
these TFT devices, they can mimic a neuron membrane potential which
leaks with a very slow time constant. Yet, by applying different
voltages they can also decay much faster. In the MemScales project
these have been used to show timescale ranges of at
least up to 8 or 9 decades (unpublished results).

To round off this quick walk
through the gardens of memory, we mention that memory functionality may
be outsourced to the external environment, in an allusion to Brook's
aphorism ``the world is its own best model''
\cite{Brooks91a}. A robot or animal does not need to memorize that a
tree stands in front of the lake as long as the tree stays in sight of
the wandering agent. But the agent has to be able to re-identify the
tree from different angles, and the agent must have attention and
evaluation mechanisms which determine when and why the tree has to be
processed again -- and those mechanisms are likewise needed in memory
management. More generally speaking, internal memory mechanisms are
interacting with, and mirroring, mechanisms of reacting to outside
world information. Timescales of memory should thus be studied in
conjuction with timescales of reactivity --- reactivity to external
world/user information or reactivity to information in memory.

In the light of all of these ramifications and complications, speaking
about ``timescales of memory'' ceases to be a straightforward
affair. This is true for artificial computing systems as well as for
brains. Upon closer inspection, the customary coarse distinction
between short-term, working, and long-term memory spreads out into a
tangle of interwoven phenomena and mechanisms, which in the human brain
are served by a multitude of physiological mechanisms
and anatomical structures (discussions in \cite{FusiWang16,
  Jaeger17}). If we neuromorphic computing researchers take our mission
to ``learn from the brain'' seriously, we should brace ourselves to
meet with extreme complexity.

\subsection{The relation between physical time constants and phenomenal timescales}

Given a physical computing system and an analytical model on the one
hand, and a formal specification of an external task on the other, how
can one bring the two together with regards to timescales? For a
discussion let us assume
\begin{itemize}
\item that the task specification includes an account of desired
  phenomenal timescales --- of change, reactivity, and memory, each
  specified in terms of task-adequate modes of progression;
\item and that the analytical model is expressed in a formalism with
  time constants.
\end{itemize} 
The engineering problem that we have to solve in this situation is to
find an abstract computational model which on the one hand meets the
phenomenal timescale demands from the task specification, and on the
other hand must be definable from the analytical system model. The
primary indication about temporal characteristics in the latter is
given in its time constants. But the analytical model does not by
itself say anything about phenomenal timescales, and the task model
will often have no time constants in it. Thus the question arises, how
do phenomenal timescales arise from physical time constants?

This is a deep and difficult issue.  In Section
\ref{subsecTimeConstants} we argued (with due caution) that time
constants might be best understood as \emph{causal} coupling strength
indicators rather than being ``temporal''. They do not directly
translate to \emph{phenomenal} speed of change, memory or
reactivity. If one wishes to think in big terms, one might say that
our problem here is to find links between the causal and the
phenomenal. Unfortunately we can only give a superficial initial
account, boiling down to a list of rather unconnected observations:

\begin{itemize}
\item A single time constant $\tau_i$ for a variable $x_i$ in a system
  of a large number of ODEs will normally not say much about
  phenomenal temporal properties of the entire system. However, when
  there is a sufficiently large subset of ODEs that are strongly
  coupled to each other and weakly to other ODEs outside this subset,
  and all of the variables in this subset have small time constants,
  one can expect that the \emph{collective dynamics} of this subsystem
  will show traits of fast timescales of change, memory, and
  reactivity. And if all time constants in this subsystem are large
  compared to ``outside'' time constants, we may expect relative
  phenomenal slowness of this subsystem. 
\item A
  set of physical system variables that collectively supports a slow
  phenomenon need not be a set of mutually densely coupled
  variables. A good example can be found in the paper by
  \citeA{Maassetal01} which is widely cited as the original reference
  for liquid state machines (LSMs). A recurrent ``reservoir'' network of
  spiking neurons solves (among other) a task with demands on
  dynamical memory length that are much slower/longer than the spiking
  dynamics. The paper does not investigate this riddle. We studied
  this paper in some depth and came to the conclusion that the long
  memory spans result from a specific synaptic adaptation mechanism
  which has slow ODE time constants (finding out about them required
  us to dig down two levels in the citation tree -- the original LSM
  paper does not document them). These slow time constants belong to
  variables whose equations are not directly coupled with each other
  but only indirectly through other, fast (spike related)
  variables. In own work where we wanted to classify slow biological
  signals with fast analog spiking neuromorphic hardware we were left,
  after much heuristic experimentation, with the impression that it
  does not really matter where, exactly, slow time constants sit in a
  recurrent neural network model, provided there are enough of
  them. This is of course only a superficial impression.
\item While the mathematical and conceptual connections between time
  constants and timescale phenomena are not very well understood,
  there is the commonplace empirical evidence that phenomenal
  timescales often \emph{do} correlate with time constants. After all,
  if one scales all time constants in an analytical system model by
  the same factor, all observable phenomenal system dynamics will
  speed up or slow down by the same factor. Smaller time constants
  will lead to faster rates of quantitative change, shorter memory
  spans, and faster reactivity. But this is a mathematical effect
  only, of little practical importance because one cannot globally
  scale all causal couplings in a physical system by the same factor.
\item While there are many ways to define from ``fast'' analytical
  variables (in that they have small time constants) computational
  meta variables that
  are phenomenally slow, the opposite seems harder and rare: namely to
  obtain fast phenomena on the basis of ``slow'' analytical
  variables. One could say 
  that if one needs fast computing, one must have fast physics. While
  this seems natural and obvious, it actually is not. In alarm
  situations, our biological brain can ``compute'' decisions in
  milliseconds, based on the integration of extreme volumes of sensor
  input. But the brain's time constants and signal travel delays are
  notoriously large compared to digital computers.  This may be a hint
  that phenomenal fastness (here: high reactivity) may emerge from
  slow physics in \emph{spatially distributed, parallel} information
  processing. A guiding role model for further analyses might be the
  picture of a lake in a mild breeze which creates a lake-wide pattern
  of smooth, low-amplitude, slowly rolling waves. If one screens this
  scene with a slowly rotating beam of vision one will record 
  fast oscillations.
\item In the idealized view that theoretical computer science has of
  digital systems, these systems have but two time constants: the very
  fast one of clock cycles and logical gate switching, and the
  infinitely slow one of perfectly non-volatile memory cell
  states. Real digital systems come amazingly close to this ideal
  picture, with memory cell techologies that have time constants so
  large that they appear as infinite for most practical matters. This
  enables digital programmers to realize any phenomenal timescale that
  lies between the clock cycle time and infinity because information
  encodings can be written to memory cells at any clock cycle time and
  accessed after whatever time has passed. In a drastic (and
  superficial) generalization to neuromorphic and other unconventional
  computing systems one might be tempted to hypothesize that
   \begin{itemize}
       \item if one has a way to physically realize \emph{fast} memory read/write operations, 
       \item and the memory mechanism time constants are {slow}, 
       \item then one should be able to computationally realize phenomena on all timescales in between. 
   \end{itemize}
\end{itemize}

We have to leave it at that for now.

\subsection{Absolute and relative timescales} \label{secAbsRel}

\emph{Online} computing tasks are specified with regards to physical
time $t$, and the generation of output signals must follow the input
within guaranteed, typically short latency. Such tasks can be served
by computing systems that operate in a cybernetic mode (see Section
\ref{secCyberneticAlgorithmic}), but also by systems which operate in
a classical algorithmic mode. In the latter case, one needs
\emph{real-time} information processing that is based on sequenced
classical sub-algorithms, which have to terminate within admissible
time windows.

Designing computing systems for such online tasks can
be challenging for a number of reasons:
\begin{itemize}
\item The formalism used for the abstract computational model must be
  expressed in physical modes of progression for the input and output
  variables, which leads to formalization and design challenges when
  intermediate processing is most naturally expressed in terms of
  atemporal logical modes of progression. An example is control and
  action selection architectures for cognitive mobile robots, where
  continuous sensor-motor control loops must be integrated with
  naturally symbolic planning routines.
\item If the underlying hardware is non-digital and only commands on
  volatile memory mechanisms, its physical time constants must be such
  that, in concert with the abstract computational model, all physical
  timescale constraints for speed of change, reactivity, and memory
  must be met.  If all one has available is a physically fast
  computing substrate but one wishes to use it for a slow task, one
  has to define slow enough meta variables for the abstract
  computational model to exploit them for the task. This was the main
  challenge in one of our projects \cite{Heetal19} where we had to
  realize a heartbeat monitoring task on an analog unclocked
  neuromorphic microchip whose native time constants were ``too fast''
  for the comparatively slow timescale of change and long memory
  durations in the incoming ECG sensor signal. If conversely the
  available hardware is too slow for the task, one would have to find
  ``faster than hardware physics'' meta variables. We lack an example
  and are not aware of systematic investigations into this
  problem. This would be a relevant and interesting project. One
  idea might be to exploit spatial patterns that change slowly in time
  but whose spatial structure can be transformed into high-frequency
  signals, as we earlier suggested in the ``waves on a lake'' metaphor.
\end{itemize}

In digital real-time systems the design of chip and computer
architectures, operating systems, and programming paradigms can become
challenging. The higher the complexity of computations that need to be
carried out in a given (real) short time window, the shorter the clock
cycle time must be. But there are limits to speeding up clock cycle
times. The trend in sub-10 nm technologies even goes rather the
opposite direction: on-chip delays become ever longer compared to the
fast clock cycles used in modern processors and systems-on-chips
\cite{Saraswat06}.

In such online processing tasks we say that the computing
system must meet \emph{absolute} timescale requirements.

A different situation occurs in \emph{offline} processing tasks, where
a data structure is inputted to a computing
system at the start of a computation run, and the result is outputted
at the end of the run. This is the classical view of executing an
algorithm in the digital-symbolic paradigm. The paradigmatic scenario
is to write the entire input on the tape of a Turing machine before it
is started. It seems that the only remaining question of interest is
how fast the computing run can be finished. In digital computing
systems, this depends only on the clock speed.

In some kinds of tasks, the input data has an intrinsic sequential
structure, and the task requires a sequential processing of the input,
where later processing stages needing access to memory traces of
previously preocessed input sectors. This is the case in many speech,
text or video processing tasks. This leads to timescale constraints if
unconventional hardware is used that only has volatile memory
subsystems. When the task demands the integration of previous input
sectors over several memory timescales, the computing system must
provide these, by a corresponding hierarchy of memory timescale,
regardless of whether they are realized through hardware or are
computationally created. The absolute real-world processing time does
not matter here except maybe for practical or economical aspects. We
then speak of the necessity to have a choice of \emph{relative
  timescales} available. The absolute processing time for such
intrinsic multiscale sequential input is bounded from \emph{above} by
the longest memory timescale required by the task: 
the machine must be run fast enough to warrant that its longest
realizable memory timespan covers the processing time between the
currently processed input sector and all previous sectors from which
memory traces are still needed.

\subsection{Section summary} \label{subsecCTEclarified}

  Summarizing the findings of our theoretical explorations, the task
  of ``computational extension of timescales'' can now be re-stated in
  more detail and with better strategic guidance. 
Our original intuitions, as stated in the Introduction, were
  quite natural and straightforward: solve the mismatch between (i) a
  few available physical timescales and (ii) task-demanded other
  timescales by finding computational ``tricks'' to ``extend'' the
  physical timescales to the needed ones.
  
But the task is not  as simple or easily stated as that:
\begin{itemize}
\item We find it insightful to study ``timescales'' on the background
  of a conception of a computational system which presents itself in
  three layers: the physical hardware system; its reflection in an
  analytical system model which attempts to model the physical system
  in a veridical manner (that is, one can specify degrees of how
  accurate or even ``true'' this system model is); and abstract
  computational models which can be defined from the analytical model
  in many different ways depending on which sort of computational
  tasks one wishes to get done with what abstract procedure.
\item There is not a uniform single concept of ``timescale''. We find
  an assortment of conceptually, physically, and mathematically
  different ideas, all of which relate to ``timescales''. This
  assortment includes the effects and phenomena which we highlighted
  in Sections \ref{subsecTimeConstants} --- \ref{subsecMemTimescales}:
  time constants reflecting strengths of causal couplings, and
  phenomenal timescales of change, reactivity, and memory. Future
  investigations will most likely reveal more kinds of phenomenal
  timescales. Further differentiations arise from considering how
  input signals relate to time (Section \ref{subsecInputtime}) and
  from distinguishing absolute from relative timescales (Section
  \ref{secAbsRel}).
\end{itemize}
  
On this background, the (too) generally stated task of ``computational
extension of timescales'' splits into a spectrum of different sorts of
conceptual, modeling and design challenges that need to be considered
separately. We have explored that terrain in the previous sections and
need not iterate our multi-faceted findings here. We wrap up by
pointing out that all of these challenges occur, in various mixes,
when one asks the two eternal questions of systems engineering:

{\bf What can a system do?} When a physical computing system is given
--- or a class of them, e.g.\ all analog spiking neural-network VLSI
microchips with a specific electronic neuron model and synaptic
connectivity options --- what computational tasks can be solved with
it? This bottom-up problem statement triggers difficult follow-up
questions, like in the following selection:
\begin{itemize}
\item Which physical phenomena seem promising for computing and how
  are they modeled in an analytical system model?
\item How can the phenomena that are made accessible through an
  analytical system model be formalized in basic computational
  models?  (In DC, these would be assembler programs calling machine
  instruction commands.)
\item What are possible input and output formats? How can task
  information be encoded in physical system input and output signals?
  What interfacing infrastructure is needed?
\item What are the phenomenal timescale characteristics (change,
  reactivity, memory) offered by the analytical model, and how can
  they be extended in a hierarchy of computational models?
\item What type of tasks can be served on the basis of a given
  analytical model and basic format of lowest-level computational
  models? Online and/or offline tasks, complexity and precision
  limits, reproducibility of ``runs'', endurance, error dignosis and
  repair, performance guarantees, costs (e.g.\ of energy, fabrication,
  maintenance)?
\end{itemize}

{\bf What system can do it?} When starting from a computational task
(or a family of tasks), the top-down engineering question is, which
physical system(s) can solve it (best)? Again, this embracing question
quickly splits into subproblems, for instance:
\begin{itemize}
\item Observing that a ``task'' is usually first expressed in informal
  terms by an end-user, how can it be formalized to enable a
  systematic analysis? In digital software engineering, many tools and
  formalisms have been developed to this end (like UML diagrams for
  practical system design and logical task specifications for task
  objective verification). This task formalization problem is the
  mirror version of the last problem in the previous list of
  bottom-up questions, and has to be solved together with it.
\item How can the phenomenal timescale requirements of the
  (formalized) task be accomodated in high-level abstract
  computational models?
\item Assuming one has a high-level computational model serving the
  task, which may be expressed in terms of a quite abstract modes of
  temporal progression $\mathfrak{t}$, how can it be ``down-compiled''
  into analytical models that use physical time $t$? And which sorts
  of analytical models are possible as ultimate basis for the
  high-level computational model?
\item If one has worked out a choice of analytical system models,
  which physical systems can realize them?
\end{itemize}

Obviously, in concrete system engineering, the bottom-up and top-down
quests will interact in design-analysis cycles. For digital hardware,
answers to all of these questions are well-known and routinely
handled. For neuromorphic and other unconventional hardware, these
questions are in general wide open --- in particular, the twin
question of which sorts of tasks can be served by what unconventional
physical systems. This question has only recently begun to be
systematically investigated. The most sophisticated answer that we are
aware of is the design framework of \cite{Zhangetal20} for a
restricted class of digital spiking neuromorphic microchips, for which
the authors specify a compilation hierarchy from formal task
specification to target hardware platforms. Another instructive
example is recent collaborative work done in our 
MemScales project \cite{Payvandetal22}, where the Self-Organizing Recurrent
Network (SORN) model of \citeA{Lazaretal09}, which previously had only
been digitally simulated, was realized on analog spiking hardware with
memristive synapses. This resulted in a hardware-software co-design
development process, where the abstract computational model was
optimized for two physically available timescales of synaptic
plasticity (represented through time constants in the analytical
model). It was also optimized for the device variability and
nonstationarity of the physical system. In the words of the
researchers, ``with the \emph{technologically plausible algorithm
  design}, we aim to optimize the hardware implementation of
algorithms by taking the hardware physics into account while
developing the algorithm'' \cite{Payvandetal22}.

\section{A zoo of computational mechanisms} \label{secZoo}

In this second part of our article we give a condensed survey of
computational mechanisms and phenomena which either directly relate to
some sort of timescale manipulation, or could be used to that end in a
natural manner, or provide useful background information to diagnose
unexpected temporal behavior. We cover almost all mechanisms that we
are currently aware of, but we are also aware that this listing is
incomplete. We roughly order our reports according to the kind of
timescale (of change, reactivity, memory, mixed and other).

\subsection{Mechanisms for managing timescales of change}

\subsubsection{Explicit training of velocity changes in temporal pattern generation}

In a variety of signal generation tasks one wishes to change the
overall speed of change of the abstract computational model when it
becomes executed by a physical system. For instance, in robot motor
control one may wish to have motor pattern generation modules that
have a velocity control input, such that  for instance a walking
gait or a pointing gesture can be generated in slower or faster
versions. In digital programming one can achieve this by letting the
program numerically solve ODEs and globally scale all their time
constants in proportion to the desired speedup / slowdown. But this
convenience will not be available in abstract computational models
defined from analog neuromorphic or other unconventional hardware.

Motor control areas in biological brains or spinal neural
\emph{central pattern generators} obviously can change the speed of
generated motor patterns. These systems are preferably modeled by RNNs
in computational neuroscience and neuro-robotics, and RNNs in various
degrees of abstraction from biology are a popular abstract model class
in analog neuromorphic computing. This raises the question how the
timescale of change (the ``velocity'') of a pattern-generating RNN can
be controlled when global scaling of all time
constants in ODEs is not possible. This question is relevant both for
understanding biological RNNs and for non-digital neurocontrollers in
robotics. It is also of theoretical interest for the 
mathematical theory of dynamical systems.

An obvious method to obtain speed-controllable RNN pattern generators
is to explicitly train them for this task in a supervised way, where
the training input consists of a slowly varying (ramp) target speed
indicator variable and the teacher output is the desired generated
pattern in different velocity versions. It is not difficult to obtain
robust frequency-controllable periodic pattern generators with this
approach (e.g.\ in \citeA{Jaeger07a}).

A potential drawback of this method is that such training data may not
be available in real-life situations, and that the speed modulation
range is defined by a well-chosen training set of input-output
combinations during the training phase, where the training examples
should preferably be sampled rather densely across the desired
velocity range. Any untrained input might lead to an undesired output
shape modulation when the trained system is later used. Furthermore we
find it not very plausible that biological motor speed control is
learnt in this way.

\subsubsection{Controlling the geometry of effective state space}

Given the potential drawbacks of the direct training approach
mentioned in the previous subsection, one would like to afford of
other methods for RNN pattern generation velocity control, which would
require training examples sampled for smaller number of teaching
velocities, and/or which would include active feedback control for
improved stability and generalization. This would also be interesting
for computational neuroscience. However, we are aware of surprisingly
little research in this direction, with existing studies apparently
being mostly confined to the analysis of low-dimensional oscillatory
dynamical systems (an indicative brief survey is in
\citeA{wyffelsetal13}).

In own previous research we tackled this problem with another approach
that aimed for generic and robust solutions and was neither  based on
specific properties of specific low-dimensional ODE models, nor on
direct training.

\begin{figure}[htb] 
\centering
\includegraphics[width=5cm]{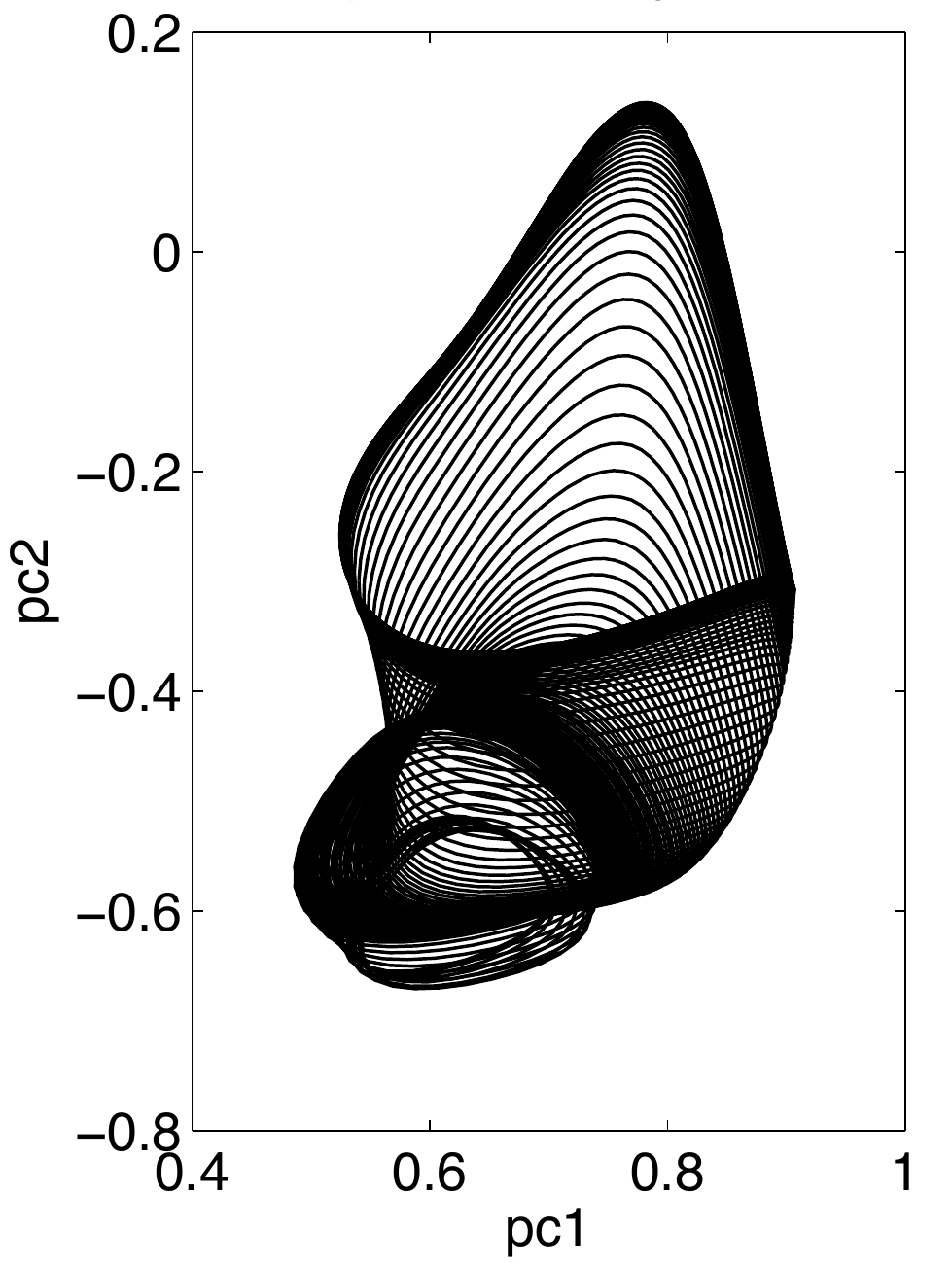}
\caption{Change of 1000-unit RNN state space region when the network
  is driven with a periodic pattern of increasing frequency (here from
  slowest to fastest with a factor of 3). The plotting axes are the
  two first principal components of the RNN states (Image taken from
  \protect\citeA{wyffelsetal13}).}
 \label{figRegionGeometry}
\end{figure}

Our work was based on the observation
that, when an RNN is externally driven with a temporal pattern, the
region in RNN state space which is visited by the entrained RNN states
systematically varies in its location and geometry when the presentation
speed of the driving signal is varied (Figure
\ref{figRegionGeometry}).

The guiding idea to exploit this phenomenon of velocity-dependent
changes in the geometry of the visited state space is to revert it: by
controlling the admissible state space region, induce velocity changes
in the generated patterns.

We pursued this line in two different ways:

\begin{description}
\item[Proportional control of RNN bias:] In \citeA{wyffelsetal13}, the
  velocity-dependent state space region was characterized simply by
  the mean of the states visited at a given frequency. This was
  translated to a bias vector $\mathbf{b}$ in a leaky integrator RNN
  upadate equation whose gain $\gamma$ was controlled by a
  proportional feedback controller that compared the frequency of the
  generated output pattern $y$ with a target signal:
  $$\mathbf{x}(n+1) = (1-\lambda)\, \mathbf{x}(n) + \lambda\,
  \tanh(\mathbf{W}_{\mbox{\scriptsize \sf rec}}\, \mathbf{x}(n) +
  \mathbf{W}_{\mbox{\scriptsize \sf feedback}} \, y(n) + \gamma \,
  \mathbf{b}).$$
  A velocity range of a factor 4 for sinewave patterns could be
  obtained with this method.
\item[Conceptor-based velocity control:] In \citeA{Jaeger14}, the
  velocity-dependent state space region was characterized by the
  principal axes of the state correlation matrix. This information was
  turned into a neural filters, one of them per task version, called
  \emph{conceptors}, which are matrices that can be inserted into the
  recurrent network state update loop and then act as a projection
  operation which nudged states outside the task-specific region back
  into it. With regards to velocity control, \citeA{Jaeger14} documents
  a simulation study where the training data consisted of only two
  sinewave samples whose frequency difference was 1 (in arbitrary
  units), from which two conceptor matrices were computed. By linear
  inter- and extrapolation between and beyond these two matrices, a
  range of new conceptors was synthesized after training which made
  the RNN oscillate in a frequency range of 5 units. 

  Conceptors have so far been mainly explored as a mechanism to train
  an RNN to generate a large number of different selectable temporal
  patterns (among them periodic patterns, chaotic patterns, and
  high-dimensional human motion patterns), yielding a lifelong
  trainable long-term memory mechanism \cite{Jaeger17}. As a bonus,
  having a conceptor in the loop has a strong stabilizing
  effect on the RNN dynamics. This effect may be particularly useful 
  in computing systems based on noisy, low-precision, or modestly
  nonstationary hardware dynamics.  
\end{description}

\subsubsection{Input-induced timescales of change in adiabatic computing processes}

Abstract and analytical models of computing systems that have internal
feedback loops (which is typically the case) can be analyzed with
formal tools from dynamical systems theory. In many cases the internal
timescales of the processing systems (often referred to as
\emph{intrinsic} or \emph{native} system timescales) will be fast
compared to the timescales of the incoming input signals. When there
is such a \emph{separation of timescales}, and when the internal
dynamics satisfies certain stability conditions, the internal
processing will ``adiabatically'' follow the slow input. In such a
scenario, the input can be regarded as a control parameter that
defines an attractor in the internal system dynamics, and due to the
timescale separation this dynamics always ``has enough time'' to
converge close to the attractor.

In the simplest case, the input defines a point attractor which moves
as slowly as the input changes, and the system dynamics tracks this
stable fixed point with a small lag. This sort of fixed-point tracking
dynamics has been postulated as a working principle for certain motor
control tasks in neuroscience \cite{Bizzietal92}, and it can be
regarded as a dynamical systems view on tracking feedback control.

The slowly modulated attractor need not be a point attractor. If it is
a periodic attractor, the slow input may, for instance, change the
geometry, offset, or frequency of the attractor. From a dynamical
systems perspective this would be the natural modeling and design
approach for \emph{central pattern generators} in neural and robotic
motor control.

To prevent confusion of terminology: We used
the word ``adiabatic'' here not in the sense of thermodynamics but to
indicate a mathematical dynamical phenomenon in slow-fast dynamical
systems that can be intuitively understood as a slowly moving target
state on an invariant manifold which ``keeps pulling'' the global state
always close to it, yielding \emph{adiabatic flows} in the limit of
infinite timescale separation \cite{Werneckeetal18}.  In biological
and robot motor control one can design tracking controllers on the
basis of this principle, which is then called \emph{equilibrium point
  control} \cite{Bizzietal92}. This usage is only distantly (if at
all) related to the understanding of adiabatic process in
thermodynamics. The latter also play a role in designing
energy-efficient digital circuitry \cite{ReddySatyamKishore08,
  NandalKumar17}, a topic that we do not further pursue here.

The idea of adiabatic tracking is natural and has been explored and
exploited many times.  However, upon closer inspection, a number of
difficulties appear.

A conceptual difficulty concerns what should be understood as the
``native timescale'' as compared to the input timescale. In the
mathematical theory of multiple timescale dynamics \cite{Kuehn15},
the main approach to formalization is to analyse ODE-based models
whose equations split into two groups, giving a ``fast'' subsystem
with small time constants and a ``slow'' subsystem with large time
constants. Timescales are here identified with time constants. The
mathematical analysis of such \emph{slow-fast systems} has been
carried to great depths, giving rise to \emph{singular
  perturbation theory}. The fast dynamics can often be approximately
understood as transient convergence toward attracting \emph{invariant
  manifolds}, and the slow dynamics as state evolution inside these,
with the current state acting as point attractor in directions
orthogonal to the manifold.  However, in computing systems the
abstract computational model may not be based on ODEs or other
formalisms that have (analogs of) time constants, and the input signal
can remain entirely un-modeled. This makes it difficult if not
inappropriate to apply the theory of slow-fast systems. In our
terminology, it would be appropriate to consider the input timescales
and the induced slow changes of the internal attractors as timescales
of change, and the transient tracking dynamics as timescales of
reactivity.

A further  difficulty is that the slow input, regarded as control
parameter in the sense of dynamical systems theory, may induce
bifurcations in the computing system. This will render it difficult to
purposely and predictably design computing systems as natively fast
systems controlled by slow input.

Another difficulty is that inputs may not be guaranteed to be
slow at all times. When there are fast changes in the input, the
tracking may lose contact with the currently followed attractor and be
pushed into another attractor's basin.

\subsubsection{Slow feature analysis}

Real-life signals $(\mathbf{p}(n))_{n \in \mathbb{N}}$, where each
$\mathbf{p}(n) \in \mathbb{R}^d$ is by itself a high-dimensional
\emph{pattern} like a video frame, a sensory input, or a neural
network state, can be described through defining different
\emph{features} that change on different timescales. By definition, a
feature is a function $f: \mathbb{R}^d \to \mathbb{R}$. Features can
be arbitrarily complex and nonlinear. Consider, for instance, a
5-minute photosafari videoclip taken from a car driving through a
savannah. For one minute, the video catches and holds the sight of a
leopard. Then a feature $f$ which recognizes the presence of the
leopard in a video frame --- i.e.\ this feature returns a value of 1
if this animal is present and 0 else --- is a \emph{slow} feature
because most of the time it is constant 0 or 1, and changes this value
only twice.

\emph{Slow feature analysis} (SFA, \citeA{Wiskott99,
  WiskottSejnowski02}) is a neural learning algorithm to automatically
identify slow features in such pattern signals. Its mathematical core
is a combination of a nonlinear expansion of the pattern signal
(this step is optional and needed if one wishes to obtain nonlinear
filters) followed by signal whitening and a final principal component
analysis to detect the principal direction in the expanded pattern
correlation space that has the slowest average temporal variation. By
repeating the last step, increasingly less slow orthogonal signals
can be identified. The method has sometimes been used in machine
learning, but its largest impact is in computational
neuroscience. Biologically plausible adaptations of the basic learning
algorithm have been developed, and SFA has been proposed as a
mechanism that trains hippocampal place cells (these cells encode
specific locations in an animal's habitat; such
locations do not move --- they are maximally slow, which is the key for
their identification with SFA). A compact summary of SFA is
\citeA{Wiskottetal11}.

\subsubsection{Slow transients}

Dynamical systems can exhibit periods of very slow transient state
progression alternating with periods of fast change. This can be due
to a variety of more or less unrelated mathematical phenomena,
for instance slow passages near saddle points, critical slowdown near
bifurcations, relaxation oscillators, gradient descent dynamics in
potential landscapes with locally different curvatures in different
directions, dynamics on center manifolds, or drift along line
attractors. We are not aware of dedicated computational exploits of
such phenomena and hint at this assortment of phenomena only to
highlight that a rigorous mathematical analysis of slowness phenomena
that one encounters in practice (maybe unexpectedly) needs a
case-by-case investigation. We point out that in computer simulations
of complex dynamical systems one sometimes finds that the observed
state evolution seems to be coming to a standstill, and may
erroneously conclude that it is approaching a fixed point attractor
--- while in fact the trajectory would have picked up speed again if
one would only have continued the simulation for a \emph{very} long
time. Specifically, in machine learning one should not prematurely
stop an iterative model optimization process when one judges that the
model accuracy has reached a plateau.

Interestingly, a similarly rich compendium of
scenarios for \emph{fast} transient behavior seems not to have been
explored.

\subsubsection{Time un-warping}

In machine learning for temporal tasks, input signals may be
\emph{time-warped}, that is they exhibit local velocity
variations. For example, a human speaker will sometimes speak faster
(when he/she is excited, for instance) than at other times; or
sometimes linger on a vocal or make a pause while thinking about how to
continue the sentence. Similar warpings occur in almost any real-life
observation stream. This is a nontrivial challenge for machine
learning algorithms both in training and exploitation because these
algorithms are ``warping-unaware'' and can become seriously
de-railed. This would happen, for instance, in a speech recognition
task where input time windows of a feedforward neural network model are
scaled to cover an entire spoken word, but the speaker is
lingering and the entire input window is filled with an ``aaaa...''.

A number of machine learning methods have been developed to cope with
this problem, especially in speech recognition. A classical signal
processing technique is to unwarp the observed signal (by super- and
subsampling) by optimizing its match against standardized reference
patterns. An obvious difficulty is to determine which standardized
reference pattern has to be used at any given time; this can be done
by dynamic programming methods, as in \citeA{Itakura75}. Hidden Markov
models, which for two decades have been the primary workhorse for
speech recognition before they became superseded by deep learning
methods, account for repetitive speech recording frames by
self-transitions of Markov states
\cite{Rabiner90}. \citeA{SunChenLee93} (also containing a short
overview of previous neural-network techniques for unwarping) propose
to make RNNs warping-invariant by making the stepsize $\Delta t$
proportional to the difference in norm between two incoming speech
vectors. This method has been adapted to reservoir computing in
\citeA{Lukoseviciusetal06}.

\subsubsection{High-frequency oscillations from coupling low-frequency
oscillators}

It it not difficult to couple two nonlinear oscillators, each having a
period $\Delta$, such that from the coupled system one can extract a
signal with period $\Delta/2$. All one has to do is to establish a
coupling that lets the two oscillators evolve with a maximal relative
phase difference, and extract a combined signal from both
oscillators. This can be generalized to obtain period shortenings for
higher than double frequencies $n/\Delta$. Higher-frequency oscillator
systems obtained from coupling lower-frequency ones could, for
instance, be used to create faster clock signals than would be
obtainable from ``slow'' oscillators available in the physical computing
system. 

The phenomenology of coupled oscillators has been studied
in breadth and depth in theoretical physics. We can imagine numerous
exploitations of purposely designed coupling schemes, but are not
aware that this option has been pursued in the neuromorphic or
unconventional computing literature.

\subsection{Mechanisms for managing timescales of
  reactivity}

\subsubsection{Increasing reactivity through arousal}

In cognitive neuroscience, the term ``arousal'' refers to a spectrum
of physiological conditions and cognitive effects that indicate a
degree of wakefulness, alertness, expectancy, vigilance, acuity of
sensory impressions, attention, speed of decision making, and the
like. It is a not universally defined, multi-causal and
multi-functional phenomenon. The term is used quite differently in
different theories of cognitive psychology and neuroscience. Giving an
overview is beyond our expertise.

In digital processing systems, a faint reflection of
arousal could be recognized in adaptive clock frequency regulation,
where the clock rate is increased at the expense of higher energy
consumption when high computational loads have to be dealt with. The
increase in energy consumption has a parallel in neurobiology,
where likewise higher arousal is connected with increased metabolic
rates. 

With regards to computing systems based on analog neuromorphic or
other unconventional physical systems, a not all too far-fetched
analogue of arousal may be seen in options to run the same physical
system at different levels of energy throughput, for instance by
increasing the overall voltage in electronic systems. Unless the
physical system and its analytical model variables respond entirely
linearly to changes in energy throughput, the physical dynamics will
undergo nonlinear changes (in the sense that phase portraits at
different energizing levels cannot be linearly transformed into each
other) and possibly bifurcate (phase portraits cannot be continuously
mapped onto each other).  Abstract computational models for such
``arousable'' physical substrates would change their information
processing characteristics in many ways, from subtle timing changes to
qualitatively different input responses. In order to exploit
``physical arousal'' to help solving computational tasks, much care
would be needed to balance desired effects (like faster reaction
times) against undesired ones (like higher error rates or increased energy consumption). We are not
aware that arousal-related mechanisms have been systematically
explored for non-digital computing tasks. A study of the respective
literature in psychology and cognitive neuroscience could become a
well of inspiration for identifying and engineering innovative
information processing models that make use
of energy-dependent nonlinear physical effects.

\subsubsection{Sensitivity control by positive and negative feedback}

While systematic investigations of arousal-related processing
modulations at the level of comprehensive computational tasks are
missing to our (limited) knowledge, excitability modulation has been
studied by theoretical neuroscientists at local levels of neural
dynamics. Relating to this tradition, \citeA{SepulchreDrionFranci19}
begin to develop an analysis of how global arousal signals sent to a
network of nonlinear processing units lead to qualitative changes of
the network dynamics, while the arousal signals are causally effective
only at the microlevel of changing efficiencies of excitatory and
inhibitory synaptic connections. The authors formulate their approach
in electrodynamical terms for abstract neuron models and give examples
from neuroscience, but also point out how their insights would
transfer to other kinds of multiscale collective systems, whether they
are computing systems (like analog circuits made from a collective of
microcircuits) or non-computing systems (like city traffic flows).

Starting from specific assumptions about the dynamics of single
neurons, they exemplify their approach of obtaining global and fast
control over the entire network's collective behavior by local
excitability modulation with three case studies. These case studies
range in network size from 2-neuron motifs to a 5-neuron model of the
crab stomatogastric ganglion (a deeply investigated model system in
neuroscience), and further to synthetic recurrent neural networks with
40 or 160 neurons. In the first two demonstrations, the global system
behavior can be analytically predicted from local excitability
modulation, while the larger network systems are explored by
simulation.  A highly didactic workout of this approach, with
suggestions for applications in neuromorphic computing, is given in
\cite{RibarSepulchre21}.

A main objective for this work was to explain (and possibly use for
engineering) how complex, collective computing systems can adapt their
information processing \emph{quickly} to changing demands. They
contrast their approach with traditional engineering solutions to
guide the behavior of multiscale systems with hierarchical control
architectures. Such classical control architectures (paradigmatic,
even an industry standard: \citeA{Albus93}) achieve a global control
objective through a cascade of control levels, where the global task
is formulated and controlled at the highest level, to become
hierarchically broken down into sub-control loops that address
increasingly local and faster subsystems. According to the authors, a
principal drawback of such architectures is that each step through the
hierarchy mandates a separation of timescales, such that the global
control is necessarily slow --- which cannot explain why animal brains
can \emph{quickly} respond to changing task demands. The approach of
Sepulchre et al.\ is not hierarchical and the global control objective
becomes immediately translated into local control. While we find this
idea compelling in principle, task-specific or complex control
objectives still remain beyond the reach of this approach.

\subsubsection{Faster reactivity through prediction}

Many online processing tasks can be regarded as optimal control tasks:
based on information that was integrated from past input, the
computing system must generate \emph{action} outputs which
optimize an expected future benefit. It is obvious that appropriate
action output can be generated faster if the system maintains a
representation of the (stochastic possibilities of the) future. This
is a generic condition which has been worked out in a variety of
fields and with many formalisms for representing stochastic futures,
estimating these representations from past input, and algorithms for
learning such estimation mechanisms from experience, and computing the
most promising action output on the basis of such future
representations. Entire engineering fields and scientific paradigms
have been shaped around this scenario, like stochastic optimal control
\cite{Stengel86}, reinforcement learning \cite{Kaelblingetal98}, or
the predictive brain paradigm in cognitive sciences \cite{Clark13} and
the closely related free energy principle for interpreting brain
function from first principles \cite{Fristonetal06}. A general
mathematical formalism for representing and learning stochastic,
input-driven processes as dynamical systems whose very states encode
the future probability distribution has been independently discovered
several times under the names Multiplicity Automata in theoretical
computer science, Observable Operator Models in machine learning and
Predictive State Representations in reinforcement learning and
robotics (unifying survey in \citeA{ThonJaeger11}). The spectrum of
background motivations, formalisms and algorithms is so wide that we
cannot attempt a more detailed overview here.

It may be less know to some readers that prediction mechanisms can
also be employed in the design of computing architectures and
microchips for embedded systems like cellular phones, multimedia
apparatuses, automotive in-car systems or body area networks. Such
systems operate under varying use cases and environmental conditions,
and their ongoing computational processes should optimize themselves
with respect to a number of cost criteria like energy efficiency or
memory usage --- and reactivity. In a design strategy called
\emph{scenario-based design}, a collection of \emph{run-time
  scenarios} (RTSs) is compiled at design time. An RTS describes a
specific composition of individual cost components which occur for a
bounded interval of time in specific operating conditions. The
algorithmic and hardware design of the embedded system then from the
outset includes methods for predicting the next RTS at use-time. This
prediction allows the system to tune its resources (e.g.\ by adapting
voltages or clock frequency) such that the overall expected
multi-dimensional cost is minimized. A recent survey of the overall
theory and practical implementations, including the design of
practical prediction functions, is presented in \citeA{Catthoor19}.

\subsection{Mechanisms for managing timescales of
  memory}\label{secManagingMemory}

\subsubsection{Effects of numerical precision for dynamical
  memory}

In recurrent neural network (RNN) models, information contained in
inputs $\mathbf{u}(\mathfrak{t})$ is transferred into neural states
$\mathbf{x}(\mathfrak{t})$ through their input laws, for example via
$\mathbf{x}(n) = \sigma(W_{\mbox{\scriptsize recurrent}} \,
\mathbf{x}(n-1) + W_{\mbox{\scriptsize input}}\, \mathbf{u}(n))$ (we
discuss this here for dimensionless discrete time $n$, but our
discussion transfers to other modes of progression like
$t[\mbox{sec}]$, $t^\emptyset$, $n\Delta[\mbox{sec}]$ in an obvious
way).  Information $\mathbf{u}(n)$ that has been infused at time
$n$ becomes amalgamated with the previous state
$\mathbf{x}(n-1)$, and this mixture is transported forward to the next
timestep where it is mixed with the next input via
\begin{eqnarray*}\mathbf{x}(n+1) & = & \sigma(W_{\mbox{\scriptsize rec}} \,
\mathbf{x}(n) + W_{\mbox{\scriptsize in}}\,
        \mathbf{u}(n+1)) \\
 &  = & \sigma(W_{\mbox{\scriptsize rec}} \,
\sigma(W_{\mbox{\scriptsize rec}} \,
\mathbf{x}(n-1) + W_{\mbox{\scriptsize in}}\,
\mathbf{u}(n)) + W_{\mbox{\scriptsize in}}\,
        \mathbf{u}(n+1)),
\end{eqnarray*}
etc. Thus a network state $\mathbf{x}(n)$ will contain traces of all
previous inputs.  This is the principle of \emph{dynamical memory}. However,
input traces thin out as time progresses, because they are
superimposed by inputs which arrive later, and furthermore they become
nonlinearly transformed at each new timestep. To make use of these
memory traces and obtain a memory signal $y(n)$ for some subsequent
processing operations, a decoding filter $y(n) =
D(\mathbf{x}(n), y(n-1))$ must be designed. This decoding filter may
have an internal state $\mathbf{d}(n)$:
\begin{eqnarray*}
  \mathbf{d}(n) & = & F(\mathbf{d}(n-1),  \mathbf{x}(n), y(n-1)),\\
  y(n) & = & D(\mathbf{d}(n)).
\end{eqnarray*}

Since dynamical neural memory is important both in neuroscience and
machine learning, a substantial literature is available where this
effect has been studied for different sorts of RNN models, tasks, and
decoding operations (a selection: \citeA{Jaeger02b, Maassetal01,
  Gangulietal08, HermansSchrauwen10, CharlesYinRozell17,
  Voelkeretal19, GononGrigoryevaOrtega20a}).

Due to the thinning-out and iterated nonlinear transformations, such
decoding filters become increasingly complex, nonlinear, and
vulnerable to noise when longer memory spans are needed. Whether a
successful decoding is possible furthermore depends on \emph{which
  aspect} of previously inserted information has to be retrieved.

An obviously favorable condition for achieving long memory spans is a
high numerical precision in the network model. The more precisely
(greater bitlength, less noise) network states are computed, the more
information inserted into them at earlier time will be recoverable. In
Turing-equivalent digital models of computing, arbitrarily high bit
precision and zero noise are possible. In non-digital
computing hardware this convenience is unavailable: abstract
computing models that can be defined from the corresponding
analytical system models will necessarily have limited precision. In
the neuromorphic field, this has spurred investigations into the
effects of limited precision in (not necessarily recurrent) neural
networks. Much of this research is motivated by 
replicating the successes of deep learning in spiking neural
networks on dedicated microchips, which often offer only bit-limited
fixed-point arithmetics (in their digital sections) or low-precision
and noisy arithmetics (in analog or memristive sections). Limited
precision affects dynamical state variables as well as synaptic
weights. These lines of research usually focus on low-precision
tolerant learning algorithms which replace or modify the
backpropagation algorithm which is a key
for deep learning. A variety of approaches have been proposed
(partly surveyed in \citeA{Indiveri15, Hashemietal17, Guo18}). The goals of
these proposals are however mostly different from ensuring long memory
spans. A systematic scrutiny of the usefulness of the proposed methods
for dynamical memory preservation remains to be done.

High numerical accuracy with its benefits for preservation of memory
traces over extended timespans is a convenience that is only available
in digital computing models. When the given hardware system is
non-digital and quantitative values of meta-variables are not
\emph{simulated} in the abstract computational model, but directly
represent physical quantities, these meta-variables in the abstract
computational model inherit the noisiness and other numerical
imprecisions from the physical system. A method to compensate to some
extent for this lack of numerical precision, and realize abstract
computations that still have reliable dynamical memory characteristics,
is a simple and generic technique which has been independently
(re-)discovered several times for different purposes under different
names: as \emph{self-predicting networks} for augmenting the
performance of reservoir computing techniques \cite{MayerBrowne04}, as
\emph{equilibration} for enabling external controllability of RNN
dynamics \cite{Jaeger10b}, as \emph{reservoir regularization} for
improved stability of neural motor controllers \cite{ReinhartSteil10},
as \emph{self-sensing networks} for making RNN training methods more
flexible \cite{SussilloAbbott12}, and as \emph{innate training} for
reliable, noise-resistant reproducible chaotic patterns in short-term
memory RNNs \cite{LajeBuonomano13}. This method was also a key enabler for
the conceptor-based training of RNNs to generate a large number of
different temporal patterns in \citeA{Jaeger14}, and for the
\emph{reservoir transfer} method proposed by \citeA{Heetal19}, where
some of the beneficial numerical qualities of a RNN simulated on a
digital computer with high precision were transferred to an RNN
implemented on an analog spiking neurochip.

This method simply re-computes the internal weights
$\mathbf{W}_{\mbox{\scriptsize \sf rec}}$ of an RNN by driving this
network with some input signal $\mathbf{u}(n)$, collecting the
resulting RNN states $\mathbf{x}(n)$ and then calculating a new weight
matrix $\mathbf{W}^\ast_{\mbox{\scriptsize \sf rec}}$ as
$$\mathbf{W}^\ast_{\mbox{\scriptsize \sf rec}} = \mbox{\sf
  argmin}_{\tilde{\mathbf{W}}}\; \sum_n \parallel \tilde{\mathbf{W}}
\mathbf{x}(n) - \mathbf{W}_{\mbox{\scriptsize \sf rec}}\,
\mathbf{x}(n) \parallel^2 + \;\alpha^2 \parallel \tilde{\mathbf{W}}
\parallel^2_{\mbox{\scriptsize \sf Fro}}.$$ This gives a
Tychonov-regularized (also known as ridge regression) version of the
original weights. In plain wording: the RNN with the new weights should
approximately replay the original state sequence with minimized
weights. The benefit is that smaller absolute weights render the
resulting network dynamics less sensitive to noise.

While in the neuromorphic computing field, research on dynamical
memory is mostly carried out in a context of RNN models, the basic
insight that numerical precision correlates with achievable memory
spans likewise applies to abstract computational models other than
RNNs. This includes digital signal processing systems and, quite
generally, any digital system that is used in online
tasks. Substantial effort has been spent in the
digital world to analyse and mitigate the effects of limited bit
precision. Like in the studies on
limited precision in neural networks, the aims of these efforts are
mostly different from ensuring long dynamical memory spans, but again
it would be worthwhile to inspect these methods and analyses with
regard to the preservation of information in successive system
states. The survey in \citeA{Menardetal19} may serve as a
particularly productive starting point because it focuses on the
propagation of rounding noise in signal processing systems.

\subsubsection{Effects of system size for dynamical memory}

This subsection can be seen as a postscriptum to the previous one,
where we pointed out that the effectively usable information encoding
capacity of a state $\mathbf{x}(n)$ grows with numerical precision and
reduction of noise. Another way to increase the capacity of states
$\mathbf{x}(n)$  to transport input information forward through time
--- the same information for longer times, or more information for the
same time --- 
is to increase the dimension of state vectors.

In unconventional computing materials with a spatiotemporal dynamics
that evolves  in a continuous space $\mathcal{S}$, states are not
vectors but functions $\mathbf{x}(n): \mathcal{S} \to \mathbb{R}$,
which are infinite-dimensional. In mathematical principle, this might
allow for unbounded information encoding capacity if the spatial
patterns could become arbitrarily finely structured. In real materials
however the local spatial pattern  variation is ultimately limited by
the discreteness of materials at the atomic level, and well before
this limit is reached, by correlations and statistical dependencies between local signals at separate points in the   substrate. Yet, it seems promising
to engineer continuously extended nonlinearly excitable material
substrates and study how abstract computational models can be
defined from them.

The theory of \emph{hyperdimensional computing} (HD) \cite{Kanerva09} can
also be mentioned as an approach to invest in system size for
precision gains. In HD, long random bitstrings are used as representations
for computational variables. Logical operations and hierarchically
nested data structures can be defined and algorithmically operated. HD
principles are being investigated for effective implementations of
brain-inspired machine learning in neuromorphic hardware
\cite{Karunaratneetal, Imanietal21}.

\subsubsection{Effects of system heterogeneity  for dynamical
  memory}

We continue the thread from the previous two subsections. Regardless
of whether states $\mathbf{x}(n)$ of a dynamical memory subsystem are
finite vectors or functions, the amount of information that can be
encoded and transported forward in time is limited by statistical
dependencies between vector components or different locations in a
continuous medium. For higher capacities of information encodings it
is thus favorable to have states whose component dynamics are only
weakly coupled in the sense of small mutual information. One way to
aim for this goal is to design memory subsystems whose state component
coupling laws are heterogeneous in some way.

In the field of reservoir computing, a diversity of proposals have
been made to design RNN matrices $W_{\mbox{\scriptsize rec}}$ and
feedback gain regimes which lead to ``rich'' dynamics. Many of these
proposals recommend specific topological RNN connection structures
(e.g.\ modular, small-world, hierarchical, or even a ring topology),
others tailor algebraic characteristics of $W_{\mbox{\scriptsize rec}}$
(in particular the eigenvalue spectrum), and yet others invoke
unsupervised learning algorithms which aim at increasing the diversity
of within-reservoir signals, sometimes in a way that is optimized for
specific input statistics \cite{Lukosevicius12}. In our
perception, reported performance gains in memory-demanding tasks have
not been spectacular. This may have to do with the fact that most
reservoir computing solutions use linear decoding functions which
essentially can take advantage only of linear decorrelational effects
of within-reservoir signals induced by such reservoir optimization
methods. More decisive benefits might be obtained from increased
statistical independence of reservoir state components, which however would
require nonlinear decoding methods to become exploited. We are not
aware of research in this direction.

Many innovations in the deep learning field rely on highly structured
RNN architectures with heterogeneous modules, some of which may come
in other formats than RNNs. With regards to enhancing dynamical memory
depth, we mention in particular the approach of adding a dedicated
memory module that is analogous to the tape of a Turing
machine \cite{Gravesetal16}. But also other heterogenous
architectures, like recurrent versions of graph neural networks
\cite{Zhouetal20}, in principle offer options for enhancing dynamical
memory. While the motivations behind graph neural networks are not
primarily focusing memory functionality, this subject would deserve
in-depth investigation similar to the analyses in reservoir computing.

\subsubsection{Effects of (non)linearity for dynamical memory}

When an abstract computational model includes a submodel which is a
dynamical system with numerical state vectors (for instance, an RNN),
one can study the effects of (non)linearity of this dynamical
subsystem on its dynamical memory properties. To this end one first
would have to define a measure for the degree of nonlinearity. One
straightforward approach would be to consider the Taylor expansion of
the state update function and define the strength of nonlinearity as
the ratio of the sum of all higher-order polynomial coefficients over
the linear coefficient. Then one could analyse the effects of
nonlinearity on dynamical memory characteristics.

This line of investigation has not been explored yet in mathematical
depth. In the reservoir computing (RC) literature we find a mostly
implicit, somewhat vague and general consensus that linear reservoirs
work best for dynamical memory. But this view may be premature, and
caused (i) by the general reliance in RC on linear decoding functions
(the trainable ``readouts'' in the RC terminology) in conjunction with
(ii) the fact that linear reservoirs are easier to analyse than
nonlinear ones. If one admits nonlinear readouts, one might get very
long dynamical memory with very nonlinear reservoirs. To see this,
consider a reservoir RNN of the kind
$\mathbf{x}(n+1) = \tanh(W_{\mbox{\scriptsize rec}}\, \mathbf{x}(n) +
W_{\mbox{\scriptsize in}} + \mathbf{u}(n+1) ).$ When
$W_{\mbox{\scriptsize rec}}$ and $W_{\mbox{\scriptsize in}}$ are
scaled to have large absolute matrix elements, such a reservoir will
develop an almost binary state sequence with almost all state
components $x_i(n)$ being close to $+1$ or $-1$. If will not be
difficult to (hand-)design $W_{\mbox{\scriptsize rec}}$ and
$W_{\mbox{\scriptsize in}}$ such that for input signals from a
specific task, certain key events of kind $A$ in the input will switch
one of the reservoir states to (say) $+1$ and leave it there until
another kind of key event $B$ switches this back to $-1$. A threshold
function readout could decode that this ``first $A$ then $B$''
sequence in the input and thus implement a specific dynamical memory
that has very long or even unbounded memory spans. Reservoir models
based on FPGAs would directly realize dynamics of this kind. Research
on discrete switching RNNs can look back on a venerable history where
such systems have been studied as Boolean networks
\cite{McCullochPitts43}, but their study within the RC paradigm
have only begun \cite{BertschingerNatschlaeger04, Verstraetenetal05b,
  LegensteinMaass07, Tanakaetal18}).

Linearity has also been found / declared as conducive for long
dynamical memory in the deep learning field, namely through long
short-term memory (LSTM) units \cite{HochreiterSchmidhuber97} and
their variants like the gated recurrent unit (GRU,
\citeA{Choetal14}). These are small neural circuits embedded in RNNs
which can be trained by backpropagation through time to store a real
number (typically between 0 and 1) with an exponential decay rate that
can be adapted through a trained control mechanism. The functional
memory core of such a neural building block is a linear
neuron. Further neurons in these circuits provide trainable write and
read functionality. These units can thus be seen as elementary working
memory devices, though this view was not the reason why LSTM units
were originally proposed (the motivation was to mitigate the vanishing
gradient problem of gradient descent optimization in neural network
training). LSTM and GRU units are a pivotal enabler for deep learning
solutions of temporal processing tasks, as witnessed for instance in
the current state-of-the-art transformer networks
\cite{Vaswanietal17, Devlinetal18} or recurrent graph neural
networks \cite{Zhouetal20} However, their mechanism and their
trainability hinges on the high-precision, noiseless numerics of
digital simulations of RNNs with rate neuron models, which renders
them not directly applicable in analog spiking neuromorphic hardware
systems. We mention them here for completeness.

\subsubsection{Line / plane attractors for long dynamical memory}

In dynamical systems theory, a line attractor
(or its higher-dimensional versions) in an
autonomous dynamical system $\dot{\mathbf{x}} = f(\mathbf{x})$ is an
attracting manifold $A$ in the system's state space on which
$\dot{\mathbf{x}} = 0$. Line attractors are non-generic objects, that
is, the slightest change of $f(\mathbf{x})$ to
$f^\ast(\mathbf{x}) + \varepsilon g(\mathbf{x})$ will destroy the
condition $\dot{\mathbf{x}} = 0$ with probability 1. However, the
system $\dot{\mathbf{x}} = f^\ast(\mathbf{x})$ will still have a
slightly deformed version $A^\ast$ of $A$ which is an attracting
manifold, and on which $\dot{\mathbf{x}}$ 
is non-zero but very slow.

\citeA{Schmidtetal21} propose a surprisingly simple method to create
such ``slow manifolds'' in discrete-time recurrent neural networks
with noisy leaky-integrator neurons and rectifier activation
functions. The method works by adding a specially designed penalty
term to the objective function given by the task. This penalty term
 singles out subsets of neurons such that the manifolds given
by state projection on these neurons become slow attracting manifolds
in the training process. The degree of slowness can be controlled for
each of the subsets. The method does not guarantee that the
created slow manifolds are attracting in all directions orthogonal to
them, but this is of no concern for their function in creating slow
memory timescales within the trained RNN dynamics. The authors
demonstrate that this set-up compares favourably with LSTM networks in
supervised learning tasks (training by stochastic gradient descent)
and unsupervised system identification tasks (training with an
expectation-minimization algorithm), where the task involves multiple
timescales. They furthermore provide analytical results that shed
light on how this method mitigates the exploding/vanishing gradient
problem in gradient-descent training. 

\subsubsection{Tuning networks close to criticality / chaos}

A popular theme in the RNN literature --- both in theoretical
neuroscience and machine learning, especially but not exclusively in
the RC field --- is that some sorts of computational performance is
optimal if the internal feedback gains in an RNN are tuned almost, but
not quite high enough to render the resulting self-excited dynamics
chaotic. This is called ``edge of chaos'' or, a little more rarely,
``edge of criticality''.  This dynamical regime is characterized by
particularly long dynamical memory spans and has been claimed to be
optimal for computational performance.

Some cautionary remarks:
\begin{itemize}
\item Chaos and criticality, two terms that are surprisingly often
  used in an undifferentiating manner within the same article, are two
  fundamentally different phenomena. Chaos refers to a specific kind
  of attractors in dynamical systems, which can appear already in
  1-dimensional discrete-time systems (iterated maps) and
  3-dimensional continuous-time systems (described by ODEs). Chaos is
  typically defined and discussed in the context of deterministic
  dynamics in continuous state spaces, and generalizations to
  stochastic or discrete-state systems need some care. Criticality, in
  contrast, is a concept of statistical physics and is properly
  defined only in the infinite-size limit of stochastic systems. The
  fact that both concepts are often confounded may be due to a
  superficial similarity: complex systems (deterministic for chaos,
  stochastic for criticality) dramatically change their qualitative
  behavior when control parameters pass a critical value, which leads
  to a \emph{bifurcation} (in the case of entering a chaotic regime)
  or to a \emph{phase transition} (for criticality). That bifurcations
  and phase transitions are different concepts in two 
  different mathematical frameworks is
  apparently not apparent to a sizable number of authors and
  reviewers.
\item In discussions of ``edge of chaos'' the context of reservoir
  computing it is often tacitly assumed that a reservoir transits
  from the dynamical mode where it has the echo state property (ESP, 
  \citeA{Jaeger01a}) directly into a chaotic regime when a specific RC
  control parameter (the spectral radius of the reservoir weight
  matrix) exceeds a critical value. This is not generally true. Often
  the reservoir passes from the ESP regime first into other attractor
  regimes (fixed point, periodic) before it further transits to chaos;
  and some reservoirs do not transit into chaos at all regardless of
  how large that control parameter is scaled. Furthermore, standard
  RNN models in RC are deterministic, and using the word
  ``edge of criticality'' to describe behavioral changes is misguided. 
\item The concept of chaos is primarily defined for autonomous
  dynamical systems without input. But computing and signal-processing
  systems (brains, reservoirs and others) are eminently input-driven
  systems.  Extensions of the mathematical concept of chaos to
  input-driven systems are nontrivial \cite{Manjunathetal12,
    ManjunathJaeger13a}, and a generally accepted general definition of
  chaos for such systems is not available. Authors writing about the
  ``edge of chaos'' are mostly unaware that the ``chaos'' they discuss
  is not (yet) mathematically well-defined.
\item Numerous papers with RNN simulation studies seem to re-confirm
  that ``computing at the edge of chaos'', or ``at the edge of
  criticality'', lead to optimal computational performance. However,
  these works almost invariably repeat variants of always the same few
  basic benchmark tasks --- tasks which happen to be just of the
  kind that support this claim. This has led to a perception that ``it
  is a well accepted fact that the computational potential of
  recurrent neural networks (RNNs) is optimized at what is called the
  edge of criticality'' (first sentence in a rather recent article in
  a top-tier journal, reworded here a little to make it unsearchable
  on the web). As antidote let us cite verbatim from another, highly
  cited, much earlier article written by more clear-sighted authors:
  ``Our experiment produced very different results, and we suggest
  that the interpretation of the original results [\emph{finding that
    complex computational tasks are best served at the ``edge of
    chaos'' }] is not correct'' \cite{MitchellHraberCrutchfield93},
  and ``... best computational power does not necessarily correspond
  to the edge of chaos'' \cite{LegensteinMaass07}. In fact,
  practicians of reservoir computing who in their lives have carefully
  optimized many a reservoir for a variety of tasks (such as the first author
  of this report) will regularly have witnessed that their optimized
  reservoirs land in a stable regime far away from any edge. It would
  be worth a socio-psychological study in scientific opinion-spreading
  to reveal why the ``edge of chaos'' myth continues to flourish.
\end{itemize}

Having deflated this myth, we nonetheless want to give a cautious
summary appraisal of a certain class of effects which have a bearing
on the study of dynamical memory:
\begin{itemize}
\item When feedback gains (in RC: the spectral radius) or related
  control parameters in recurrent, parallel, high-dimensional
  information processing systems (like reservoir RNNs or cellular
  automata) are scaled up toward, but not into a dynamical regime
  where self-excitation would override the entraining effects of input
  signals,
\item the state trajectories of the processing system begin to develop
  autocorrelations (or other nonlinearly or statistically defined)
  dependency measures across increasingly long timespans, indicating
  an impending transition from the usually observed exponential decay
  of these memory curves to only power-law decay profiles (so-called
  ``fat tails'' or ``1/f spectra''),
\item which in turn may be beneficial in temporal tasks that require
  the integration of input information over several timescales, since
  memory curves with fat tails indicate that input information decays
  slower than exponentially.
\end{itemize}

A conceptually clean, mathematically correct, and unifying analysis of
such effects remains to be done.

\subsubsection{Oscillation-based dynamical memory}

A possibility to encode information in a dynamical memory subsystem
for long memory spans is to
\begin{itemize}
\item design the memory subsystem as a collective of uncoupled
  oscillators with different frequencies, 
\item encode input information at initial time by letting it modulate
  the relative phases of the oscillators,
\item decode a desired output by a detector which is sensitive to
  these relative phases.
\end{itemize}
We are not aware of studies of such memory subsystems in the
literature, but can report from own simulation experiments in a
reservoir computing setting \cite{Jaeger12}. The task mimics an
experimental design which is often used in cognitive psychology to
study working memory in animals and humans. We used a version of this
type of task that had been used before for learnability demonstrations
of long-term memory spans in the deep learning field
\cite{MartensSutskever11}.

This version of the task was specified as follows.  At the beginning
of each experiment, a binary 2-dimensional, 5-timestep pattern (of
which there are 32 different ones) is presented to the network, needing 5
timesteps. Then, for a period of duration $T_0 - 1$, a constant
distractor input is given. After this, i.e. at time $T_0 + 5$, a cue
signal is given as a spike input, after which in the final 5 time
steps, the memory pattern has to be reproduced in the output units.

Here we report one of several variations of how this task was solved
with RC methods \cite{Jaeger12}. For a memory span of $T_0 = 1000$
(measured in discrete network updates $n$) we designed a special
reservoir weight matrix which realized a reservoir dynamics of 30
isolated oscillators with randomly chosen frequencies, and a ``bulk''
part of the reservoir which was randomly connected in the spirit of
reservoir computing, and driven by the external input and the
oscillators within the reservoir. All 32 input patterns could be
retrieved without error with a linear decoder trained by linear
regression, using a reservoir size of 500 neurons (including the
oscillator sub-reservoirs).

The presence of the oscillators within the reservoir, as well as the
random ``bulk'' part of the reservoir, were crucial to solve this
task.  Without oscillator sub-reservoirs, a reservoir size of 2000
neurons was needed to retrieve the 32 patterns after the much shorter
waiting time $T_0 = 200$. The random ``bulk'' part of the reservoir
presumably was needed to aid the decoder by supplying a
high-dimensional nonlinear expansion of the oscillator signals which
carried the input information through time.

We also report that a (much) more difficult version of this task with
input pattern of length 10 and $T_0 = 200$, which could not be solved with
the backpropagation-based RNN training methods available in 2011, as
reported by \citeA{MartensSutskever11}. Using the oscillator-based
reservoir approach, it could be perfectly solved for $T_0 = 300$ with
a 2000-neuron reservoir, requiring merely two minutes training time
on a 2.9 GHz, 8 GB RAM PC without GPU extension with un-optimized
Matlab code. 

Oscillator-based approaches for dynamical memory subsystems may hold
promises in neuromorphic / physical computing, because oscillators can
be realized in hardware from fast devices with small time constants,
yet the oscillations persist stably for long times. 

\subsubsection{Attractor-based working memory}\label{subsecAttractorWM}

While there is no universally defined and shared terminology, we
recall that the concept of working memory (as opposed to other sorts
of short-term memory) usually implies that
\begin{itemize}
  \item working memory systems can store only a limited, small number of
  separate memory items;
\item memory items are ``stored'' and ``recalled'' by external control
  mechanisms;
\item once stored, memory items can persist in memory for extended and
  potentially unbounded timespans, which however may require special
  mechanisms of attention or rehearsal;

\item after retrieval, the item is deleted from the working memory,
  freeing capacity to store another item. In neural systems it is
  typically implied that the storage of an item is a dynamical
  phenomenon which does not lead to persistent structural (synaptic
  long-term memory) changes.
\end{itemize}
In digital computing systems, designing working memory mechanisms
poses no principled difficulties with respect to hardware or control
architectures --- except that the notorious bottleneck in von-Neumann
architectures is a major obstacle for computational speed and energy
efficiency --- and we will not further discuss it. In contrast, it is
not easy to obtain working memory functionality in neural processing
models, neither in neuroscience modeling nor in artificial neural network
engineering. The difficulties are twofold: firstly, neural processing
systems typically do not afford of non-volatile memory elements in
which the information of a memory item can be safely encoded and
preserved until recall and deletion, and secondly, the requisite
mechanisms of evaluation (what should be put into working memory
when), encoding, rehearsal, decoding and resetting are complex and
need involved control mechanisms outside the item-storing subsystem
proper. Artificial neural networks in machine learning are
\emph{trained}, and it is hard to train these control mechanisms. They
need to be pre-designed at least partly, as in the spectacular study
of designing a ``neural Turing machine'' \cite{Gravesetal14,
  Gravesetal16} and in our RC-based model of a hierarchically
structured working memory \cite{PascanuJaeger10}.

The literature in cognitive neuroscience on working memory
is extensive and we can only point out three surveys of
\citeA{Durstewitzetal00}, \citeA{Baddeley03} and \citeA{FusiWang16}.

The most natural candidate for representing and stable storing of
discrete memory items in dynamical (neural) systems are
\emph{attractors}. This connects the study of working memory to a
deep, ancient, and unresolved challenge in neuroscience and
neural-network based machine learning, namely the problem of how
discrete, symbolic ``knowledge'' can be neurally represented and
processed. This theme can be traced back at least to the notion of cell
assemblies formulated by \citeA{Hebb49}, reached a first culmination in
the formal analysis of associative memories \cite{Palm80, Hopfield82,
  Amitetal85a}, and has since then diversified into a range of
increasingly complex models of interacting (partial) neural attractors
and associated ``attractor-like'' phenomena \cite{YaoFreeman90,
  Tsuda00, Rabinovich08, SussilloBarak13, Schoener19}. But other
geometrical objects have also been selected as representatives of
concepts, for instance solitons and waves \cite{LinsSchoener14} or
algebraically defined regions in state space
\cite{Tinoetal98,Jaeger17}, and many other cognitive entities besides
concepts have been modeled by qualitative phenomena in dynamical
systems \cite{GelderPort95, Besoldetal17}. 

With regards to the topic of this article, attractors are of
fundamental interest because they embody two timescales of change ---
namely the fast timescale of the ongoing state evolution in
periodic and chaotic attractors, and the unlimited timescale of stably
being ``locked'' in the attractor --- and a timescale of reactivity,
namely the (typically exponential) transient convergence toward the
fully developed attractor dynamics. For purposes of supporting working
memory functionality, one has to design or train memory subsystems
which host many attractors, one for each possible discrete memory
item. The ``storing'' task translates to control
mechanisms which push the memory subsystem into the basin of
attraction of the attractor that represents the targeted memory
item.

There is a virtually unlimited freedom of designing dynamical (neural)
systems which host many attractors and have control mechanisms to
steer the overall system trajectory into specific attractors. The
classical model of human working memory, the \emph{phonological loop}
model of \citeA{BaddeleyHitch74, Baddeley83}, employs (in our
terminology) a cyclic attractor. We cannot attempt a comprehensive
overview on attractor-based working memory and instead point out to
own research in this domain. In \citeA{JaegerEck08} a rather small
reservoir system of 100 neurons is trained in a way that it hosts in
the order of $10^{7}$ (!)  different cyclic attractors, each
representing a short musical melody motif, where the system is steered
into the desired attractor by entrainment to a short cue sequence. 
In \citeA{PascanuJaeger10} a neural
reservoir is steered into different processing modes, each of them
carrying out a text prediction task with different underlying grammars
for each mode, where the switches between modes is triggered by
opening and closing brackets that appear in the text. The processing
is locked in the respective mode by switching external control neurons
in a hierarchy of on-off states. In \citeA{Jaeger17} \emph{conceptors}
are used as external control mechanism to switch an RNN into different
trained attractors, which in a range of simulation experiments
generated simple periodic signals, chaotic attractor dynamics, or
human motion patterns. --- These three studies are very different from
each other, indicating the richness of design options for
attractor-based working memory.

\subsubsection{HiPPO}

\citeA{GuDaoetal20} propose a computational method which explicitly
creates dynamical memory traces of an input signal in state vectors of
RNNs. This method, called HiPPO (from ``High-order Polynomial
Projection Operators'') can be outlined -- with some adaptation of the
original notation -- as follows:
\begin{itemize}
\item First one defines a time-indexed family of probability measures
  $\mu(t)$ with finite (but possibly time-varying) support $S(t)$ on
  the past input history up to the present time $t$.
  \item On the support of this probability measure one considers the
    linear function space $\mathcal{G}$ spanned by polynomials up to some order
    $N$.
  \item Within $\mathcal{G}$ one identifies, for every time $t$, the
    function $g^{(t)} \in \mathcal{G}$ which optimally approximates the input
    history in the sense of minimal distance to the input during $S(t)$,
    where the distance metric is the $\mu$-weighted inner product
    $\int_{S(t)} g^{(t)}(\tau)\, f(\tau) \,d\mu(\tau)$ of  $g^{(t)}$ with
    the input signal $f(t)$.
  \item With respect to a diligently chosen basis of this function
    space one represents $g^{(t)}$ as the $N$-dimensional 
    vector $c(t)$ of the projection coefficients of  $g^{(t)}$ on the chosen
    basis functions.
  \item For several natural and practically relevant types of $\mu$,
    closed-form solutions for $c(t)$ are derived which can be
    mathematically expressed and effectively computed in several
    formats, including one format that expresses $c(t)$ as the
    solution of the linear ODE $\dot{c} = A(t)c(t) + B(t)f(t)$ for
    $A(t) \in \mathbb{R}^{N \times N}, B(t) \in \mathbb{R}^N$, with
    explicit update formulas for $A(t) \to A(t+\Delta)$ and
    $B(t) \to B(t+\Delta)$, where $\Delta$ is the sampling interval. This admits an interpretation of $c(t)$ as
    the state vector of an $N$-dimensional linear RNN with
    time-varying synaptic recurrent and input weights.
  \end{itemize}

  The probability measures $\mu(t)$ over some segment $S(t)$ of the
  past input history provide what one could call the ``relevance
  weighting'' of past inputs for the target output at time $t$. The
  most notable, fully analysed type of such   $\mu(t)$ in
  \citeA{GuDaoetal20} concerns scenarios where the neural network is
  started at time $t_0 = 0$ and run indefinitely long. At time $t$ the
  probability measure   $\mu(t)$ is the uniform measure on the
  interval $[0, t]$. This means that all previously entered input is
  considered equally important for the current output generation at
  time $t$. 

  In follow-up work, remaining numerical instability and computational
  efficiency problems have been solved \cite{Guetal21}.

  The performance of this way of designing RNNs for tasks with
  extremely long (up to 16K update steps) relevant memory spans is
  demonstrated in several benchmarks, among them the extremely hard
  \emph{Path-X} binary decision problem, which was never mastered before better
  than random guessing, for which HiPPO-based networks achieved an
  88\% accuracy. This and other impressive gains over previous RNN
  learning methods have created a vivid interest in the deep learning
  community.

\subsubsection{Tapped delay lines}

In mathematical abstraction, a (discrete-time) tapped delay line is a
memory element which at time $n$ stores a moving window
$(\mathbf{u}(n-L+1), \ldots, \mathbf{u}(n))$ from an input signal
$(\mathbf{u}(n))_{n \in \mathbb{N}}$, together with a readout
mechanism which can address and retrieve each of the
$\mathbf{u}(n-L+1), \ldots, \mathbf{u}(n)$ at time
$n$. Continuous-time tapped delay lines are defined in an analog
fashion. Tapped delay lines can thus be regarded as an elementary and
at the same time powerful type of working memory which can serve a
wide range of functions in temporal processing tasks.

Discrete-time tapped delay lines can obviously be programmed in
digital general-purpose computers, though with a considerable
computational overhead. For higher efficiency and faster access,
another, hardware-based approach in digital systems is the use of
clocked synchronous registers where the clock period is varied. Many
technologies have been proposed for such digital register-based
delays.  However, this hinges on the availability of a clock signal
which is easily configured in periods that vary over a wide range of
scales. Despite many propositions for analog circuits which directly
implement a tuneable analog delay (for instance for use in
high-frequency channel equalization \cite{BuckwalterHajimiri04}), it
remains a largely unsolved problem in practical large circuits which
are required for a full system-on-chip.  The trend is rather to go for
a few well-defined clock periods that are hard-coded in the clock
generation hardware. 

Tapped delay line and related functionalities are necessary for a wide
range of cognitive tasks that have been experimentally studied in
cognitive neuroscience. \citeA{MaukBuonomano04} give a survey, and
\citeA{BuonomanoMerzenich95} and \citeA{Buonomano00} propose specific
(spiking) neural circuits that may support these functionalities in
the human brain. These circuits have a reservoir computing flavor in
that they exploit the activation propagation in essentially random
RNNs. Neural delay lines have also been modeled through traveling
solitons in neural field models of cortical processing \cite{Fardetal15}.

In the field of reservoir computing, a (by now) classical
demonstration of echo state networks is to train them as tapped delay
lines \cite{Jaeger02b}. This has led to a rather extensive literature
on the \emph{memory capacity} of RNNs, which is essentially defined as
the maximal window length $L$ achievable with a given RNN trained as
tapped delay line for i.i.d.\ input signals and using a linear readout
operation (for example in \citeA{Gangulietal08,
  HermansSchrauwen10, CharlesYinRozell17, Voelkeretal19,
  GononGrigoryevaOrtega20a}). 

Physical realizations of delay lines in a reservoir
computing setting have been variously studied, for instance in optical
computing \cite{Appeltantetal11} or using traveling spin waves
\cite{WattKostylevetal21}. Reservoir-based delay lines have also
been proposed for analog delay circuits in  the electronic signal
processing field \cite{BaiYi18}.

\subsection{Mechanisms with mixed impacts}

\subsubsection{Frequency filtering and smoothing}

In linear signal processing, numerous design principles for frequency
filtering are known. Such filters attenuate
specific frequency components in their input signals and let the other
components pass. One speaks of low-pass / high-pass filters if the
high / low frequency components are attenuated, and of band-pass
filters if frequencies lower or higher than a specified range of
passed frequencies are attenuated.

When a signal becomes low-pass filtered, its timescales of change
become longer: the filtered signal changes more slowly. This also
changes reactivity: the filtered signal does not react to
high-frequency components in the input anymore, and may react to slow
input components only with a lag (if the filter is \emph{causal}, i.e.\ the
current output is determined on the basis only of past inputs, not of
future ones).  Finally, memory characteristics change too: filtered
signals retain slower input components for longer and forget faster
components more quickly or immediately. This account of altering memory
characteristics however is limited to ``information'' that is encoded
in frequency components. 

The effects and uses of high-pass filtering are not directly
complementary to the effects of low-pass filtering. While low-pass
filtering effectively makes the signal change more slowly, high-pass
filtering does not make it change faster. A common exploit of
high-pass filtering is the removal of baseline drift from empirical
measurement signals.

The mathematical theory of linear filters offers a host of designs to
obtain frequency filters that optimize specific quality criteria that
are mostly formulated in the frequency domain. Such mathematically
defined filters can be perfectly realized (up to sampling effects) in
digital signal processing systems by numerical computations. Analog
processing hardware however can only practically realize a limited
choice of filter designs and may need capacitors or inductances whose
area footprint in neuromorphic microchips may be
prohibitive. Specifically, applications of analog on-chip low-pass
filters for ``slowing down'' computations with highly miniaturized
neuromorphic microchips are limited.

One of the simplest low-pass filters is the \emph{exponential
  smoother}, which in its discrete-time version computes the output
signal $y(t)$ iteratively from an input signal $s(t)$ with
\emph{leaking rate} $\lambda$ by
$$y(t+\Delta) = (1- \Delta \,\lambda) \, y(t) + \Delta \, \lambda \,
s(t),$$
where $0 \leq \Delta \, \lambda \leq 1$. In RNN theory, this
filter is related to \emph{leaky integrator neurons} in that the
continuous-time RNN equation
$$\tau \, \dot{\mathbf{x}} = - \mathbf{x} + f(\mathbf{x},
\mathbf{u}),$$ when integrated with the Euler approximation and
stepsize $\Delta$, becomes
$$\mathbf{x}(t+\Delta) = (1-\Delta/\tau)\,\mathbf{x}(t) + \Delta/\tau
\,  f(\mathbf{x}(t),\mathbf{u}(t+\Delta)),$$
hence the network states $\mathbf{x}$ can be regarded as
exponentially smoothed versions of the ``input'' signal
$f(\mathbf{x}(t),\mathbf{u}(t+\Delta))$ with leaking rate $1/\tau$. 

Leaky integrator RNNs are often used to design layered RNN
architectures where neurons in the bottom layer have a large leaking
rate (i.e.\ it is ``fast'') and higher layers have increasingly
smaller ones (i.e.\ they are increasingly slower). Input signals are
typically fed to the bottom layer only, output signals are extracted
from the top layer, or the bottom layer, or all layers, depending on
the task. The guiding idea for such architectures is to mimic
cognitive processing hierarchy of (sensor) signals, where the bottom
layer provides the peripheral sensory interface (in some architectures
also the motor command output interface) and increasingly higher
layers process increasingly more abstract / comprehensive / complex
``conceptual'' representations. Letting higher layers operate more
slowly allows their states to develop ``conceptual'' representations
of information in the input stream that is integrated over longer
memory timespans, while lower layers are devoted to extracting faster
and more short-lived (and in spatially organized input, also
smaller-scale) input detail. This basic idea unfolds into a
bewildering spectrum of quite different designs for ``intelligent''
information processing tasks.  A major design decision is whether
these layers are only coupled unidirectionally bottom-up, or whether
also top-down couplings are included; and another is whether these
systems are trained in a supervised way (for example for
classification, prediction, or motor control), or in an unsupervised
or reinforcement learning paradigm (for concept discovery or behavior
optimization in autonomous robotic agents). We cannot attempt a survey
here and only arbitrarily pinpoint the lifetime works of Jun Tani who
studied such layered, timescale-spreading RNN architectures for many
purposes (e.g.\ robot control architectures \cite{NishimotoTani09} or
unsupervised gesture recognition \cite{Jungetal15}). The benefits of a
spread of time constants of membrane potential leaking for learning
real-world tasks that have several relevant memory timescales has also been
documented, from a computational neuroscience angle, in simulated
spiking neural networks, and it was found that optimized time constant
distributions match empirically found distributions in biological
brains \cite{PerezNievesetal21}.

\subsubsection{Event-based computing}

In event-based computing models, local states remain unchanged until
an incoming event triggers an update.  Since between events
local states remain unchanged, event-based computing offers a direct
opportunity for realizing slow or fast timescales of all our three sorts.

In a computing scheme with a mathematically pure event-based character
(that is, the computational flowchart would contain no real-time
timing conditions; state updates occur instantaneously; event signals
are transmitted with zero delay; and whether a local processing
subsystem updates its state only depends on whether  enabling
events have arrived), such slow-downs or speed-ups would be controlled
exclusively by the frequency of input events. However, real computing
systems do have nonzero physical times for state updates, signal
transmission delays, and analog systems will have volatile states that
do not remain unchanged between incoming events. Here the achievable
timescale changes are limited by a variety of temporal interactions
between and decay processes within local states and their
updates. This mandates careful designs. To this end, modern (digital)
microchip and computer architectures and real-time operating systems
include Reliability, Availability, and Serviceability (RAS) modules
which monitor ongoing operations and control the current processing
speed 
\cite{Rodopoulosetal15, Noltsisetal17}.

\section{Take-home messages and outlook}

In this report we inspected the challenge of ``computational extension
of physically obtained timescales'' in some detail. Let us summarize
our main findings:

\begin{itemize}
\item The original problem statement in the title of this report is
  very much underspecified. In order to relate the two keywords
  ``computational'' and ``physical'' to each other in a productive and
  insightful way, we worked out the existing \textbf{theoretical framework for
  physics-based computing} of \citeA{StepneyRasmussenAmos18} in much
  greater detail. In our view, an analytical model of the physical
  computing system is a necessary interface model between the material
  hardware and abstract computational, ``algorithmic''
  models:
  \begin{itemize}
  \item ``computational'' manipulations of timescales are defined and
    developed for the abstract computational model,
  \item and the are connected to the physical system by formal
    definition operations from the analytical model.  
  \end{itemize}
\item One hardware basis can support many different abstract
  computational models. However, when one engineers a combination of a
  physical hardware system and an abstract computational model
  \textbf{for a given task}, one has to ensure that the abstract meta
  variables can be formally defined from the analytical system
  model. In view of the enormous freedom in formulating abstract
  computational models --- a blessing and a challenge --- finding a
  solution to the combined hardware/computational model problem will
  likely result in iterative optimization cycles. A further difficulty
  is that analytical models are always approximate, which may mandate
  expensive prototype fabrication and experimental characterization. 
\item \textbf{``Timescales'' is not a uniformly defined concept.} In
  order to carry out meaningful studies and do insightful engineering,  
  \begin{itemize}
  \item one must have a clear view of \textbf{how one formalizes
      temporal progression} in the first place, with an important
    distinction between modes of progression that are tied to external
    physical time and can be measured in seconds, versus abstracted
    mathematical concepts of serial order that are dimensionless ---
    abstract computational models can use both sorts but the
    analytical system model is tied to physical time;
  \item and one must be clearsighted about the \textbf{phenomenal
      aspects of ``timescales''} that one is speaking about; we
    distinguished the aspects of (metric) change, reactivity, and
    memory.  
  \end{itemize}
  We furthermore pointed out that ``timescales'' \emph{phenomena} are
  only indirectly connected to the \emph{causal} influences that
  become expressed in the time constants of the analytical physical
  model.
\item Decades of research in computational neuroscience, dynamical
  systems, theoretical physics, machine learning / neural networks and
  other fields have generated \textbf{a wealth of formal,
    algorithmical, and heuristic-practical techniques} for analysing,
  transforming, and exploiting timescale phenomena. We collected a
  ``zoo'' of 22 species but surely missed many more.  Each of them
  would need a more in-depth scrutiny than what we could deliver here
  and for each of them practical ``tricks of the trade'' would
  be welcome.
\end{itemize}

The deeper we delved into our subject matter, the more clearly did we
perceive that we are only at the beginning of a long journey. Before
we state what we see as important future work, let us share \textbf{an
encouraging thought}:
\begin{itemize}
\item Perceiving the enormous degrees of freedom in designing
  computational models for a given hardware system, and the wealth of
  already well-studied computational mechanisms that relate to
  timescales, we find that one can (almost) always find a solution for
  a given task which frees the algorithmic model from the causal time
  constants of the physical infrastructure.
\end{itemize}

At present we still lack recognition, routine and recipes, and
solving computational tasks that involve timescale challenges on
non-digital hardware still requires heuristic trial-and-error
experimentation. Only  \textbf{future research} will bring us closer
to a point where we can swiftly produce hardware engineering +
computational model solutions to given tasks. In particular, we think
that further work on the following subjects will bring advances:
\begin{itemize}
\item \textbf{Extend the repertoire of phenomenally defined 
  timescales} beyond the three sorts that we registered in this
  report. For instance, we could imagine to introduce timescales of
  duration (how long does a process take), timescales of prediction,
  or timescales of information collection and integration.
\item \textbf{More formal modes of progression} beyond the small
  assortment that we mentioned in our report would add more options
  for the design of abstract computational models. For instance, one
  could take a closer look at the classical temporal relations
  proposed by \citeA{Allen91} in AI knowledge representation, or
  investigate how formalisms of temporal modal logic \cite{Garson14}
  can be used for algorithm design.
\item We need a systematic study of how and when meta variables with a
  specific mode of progression can be defined from other meta or
  analytical variables that have other modes of progression. It would
  be desirable to derive \textbf{a formal hierarchy of modes of
    progression}, such that modes higher in the hierarchy can be
  defined with specified transformations from modes lower in the
  hierarchy.
\item Finally, \textbf{time and space} are connected both in physical
  and abstract systems. Our report focused entirely on timescales. But
  physical systems are also organized on several spatial scales
  (metric or more generally topological), and computing tasks often
  have some kind of abstract spatial/topological organization in their
  data or task specification. There are many connections between time
  and space --- starting from the elementary fact that motion needs
  both to be defined. A complete study
  of timescales must ultimately become connected to a study of spatial
  scales.
\end{itemize}

\paragraph{Acknowledgements} This article is an extended version of a
deliverable written for the EU H2020 project ``MemScales'' (grant
number 871371 \url{https://memscales.eu}), which in turn is a
follow-up to the EU project ``Neural computing architectures in
advanced monolithic 3D-VLSI nano-technologies'' (\href{www.researchgate.net/project/NeuRAM3-NEUral-computing-aRchitectures-in-Advanced-Monolithic-3D-VLSI-nano-technologies}{NeuRAM3}). Funding
received from these projects for the groups of the authors is
gratefully appreciated, as well as the opportunities and productive
challenges afforded by the interdisciplinary collaboration within the
MemScales consortium in general. We would also like to thank all the MemScales colleagues for the many fruitful and enriching technical discussions, and Giacomo Indiveri and Bernab\'{e} Linares-Barranco in particular. and with Giacomo Indiveri and
Bernab\'{e} Linares-Barranco in particular. Furthermore, substantial
ideas that were incorporated in this article took shape in the work of
the first author within the EU H2020 ITN ``Post-Digital'' (grant
number 860360 \url{https://postdigital.astonphotonics.uk}) which is
likewise gratefully acknowledged.

\bibliographystyle{apacite}
\bibliography{./references.bib}

\end{document}